\documentclass{aa}  

\usepackage{graphicx}
\usepackage{txfonts}
\usepackage[colorlinks=true, linkcolor=red, citecolor=blue]{hyperref}

\begin{document}

   \title{Testing the models of X-ray driven photoevaporation with accreting stars in the Orion Nebula Cluster}
   \titlerunning{Testing the models of X-ray driven photoevaporation with accreting stars in the Orion Nebula Cluster}
   \author{S. Flaischlen\inst{1}
                        \and
          T. Preibisch\inst{1}
                        \and
                  C.F. Manara\inst{2}
                        \and
                  B. Ercolano\inst{1}}
   \institute{Universitäts-Sternwarte, Ludwig-Maximilians-Universität München, 
                        Scheinerstraße 1, 81679 München, Germany
         \and
             European Southern Observatory, 
                         Karl-Schwarzschild-Straße 2, 85748 Garching bei München, Germany
             }

   \date{Received October 23, 2020; accepted February 1, 2021}

 
  \abstract
   {Recent works highlight the importance of stellar X-rays on the evolution of the circumstellar disks of young stellar objects, especially for disk photoevaporation.

        }
   {
        A signature of this process may be seen in the so far tentatively observed dependence of stellar accretion rates on X-ray luminosities.
        According to models of X-ray driven photoevaporation, stars with higher X-ray luminosities should show lower accretion rates, on average, 
        in a sample with similar masses and ages.
        }
   {
        To this aim, we have analyzed X-ray properties of young stars in the Orion Nebula Cluster determined with \textit{Chandra} during the COUP 
        observation as well as accretion data obtained from the photometric catalog of the \textit{HST Treasury Program}. With these data, we have performed a 
        statistical analysis of the relation between X-ray activity and accretion rates using partial linear regression analysis.

        }
   {
        The initial anticorrelation found with a sample of $332$ young stars is considerably weaker compared to previous studies. 
However, excluding flaring activity or limiting the X-ray luminosity to the 
soft band ($0.5 - 2.0~\mathrm{keV}$) leads to a stronger anticorrelation, which is statistically more significant. 
Furthermore, we have found a weak positive correlation between the higher component of the plasma temperature gained in 
the X-ray spectral fitting and the accretion rates, indicating that the hardness of the X-ray spectra may influence
the accretion process.
        }
   {
        There is evidence for a weak anticorrelation, as predicted by theoretical models, suggesting that X-ray photoevaporation modulates the 
        accretion rate through the inner disk at late stages of disk evolution, leading to a phase of photoevaporation-starved accretion.}


   \maketitle
%
\section{Introduction}

The interaction between young stellar objects (YSOs) and their circumstellar disks is of fundamental 
importance for an understanding of planet formation and migration \citep[e.g.,][]{Ercolano2017}. During their 
evolution, many YSOs are surrounded by circumstellar disks, from which material is accreted onto the 
protostar. This process is thought to be controlled by stellar magnetic field lines connecting the disk with the 
protostellar surface, leading to a complicated system of magnetic ``accretion funnels'' that channel the 
accretion flow \citep{Hartmann2016}. The infalling material generates shocks at the protostellar photosphere, giving rise 
to strong optical emission lines (in particular H$\alpha$) and optical and ultraviolet excess emission, from 
which the accretion luminosity $L_\mathrm{acc}$ can be inferred with spectroscopy \citep[e.g.,][]{Alcala2017} 
or photometry \citep[e.g.,][]{Venuti2014}. Together with the stellar parameters, the accretion rate 
$\dot{M}_\mathrm{acc}$ can be obtained from $L_\mathrm{acc}$ via $\dot{M}_\mathrm{acc} = L_\mathrm{acc} R / (0.8\,G\,M)$, where 
$R$ and $M$ are the stellar radius and mass, respectively, while the factor $0.8$ originates from 
the assumption that the infall stems from the magnetically truncated disk with a typical radius of 
$\sim 5$ stellar radii \citep{Hartmann2016}. 

Several observational studies support the expectation that the accretion rate is higher for more massive 
YSOs \citep[e.g.,][]{Muzerolle2003, Natta2006, Telleschi2007, Manara2012, Alcala2014, Alcala2017}. However, the observed spread of up to two orders 
of magnitude in accretion rates for stars with similar masses dilutes the $\dot{M}_\mathrm{acc}-M$ relation 
\citep[e.g.,][]{Hartmann2016, Manara2017, Alcala2017}. Since the disk material gets dispersed over time, 
the accretion rate also decreases with isochronal age \citep[e.g.,][]{Hartmann1998}. However, there is also evidence for stars with high accretion rates at old ages \citep[e.g.,][]{Venuti2019}.

The dispersal of the disk material is caused by strong disk winds. Observational evidence and 
simulations suggest that irradiation from the central star is an
important driving mechanism of these disk winds \citep{Alexander2014}. In particular, models predict X-ray emission 
to be very efficient in photo-evaporating the circumstellar material \citep{Ercolano2008b, Ercolano2008a, 
Ercolano2009, Owen2010, Owen2011, Owen2012, Giovanni2019, Woelfer2019}, affecting the accretion rate \citep[e.g.,][]{Drake2009} as well as planet 
formation and migration \citep[e.g.,][]{Ercolano2015, Ercolano2017, Jennings2018, Monsch2019}.
Compared to main sequence stars, YSOs show highly elevated X-ray activity through all evolutionary 
stages \citep{Feigelson1999, Preibisch2005, Preibisch2005b}.
Observations \citep[e.g.,][]{Preibisch2005,Telleschi2007} have shown that the X-ray luminosity 
of YSOs increases with increasing mass, analogously to the accretion rates. 

The \textit{Chandra Orion Ultradeep Project} (COUP) produced the largest available data set 
of X-ray emitting YSOs in a star forming region \citep{Getman2005b}. The analysis of these data showed 
that, on average, accreting YSOs have lower X-ray luminosities than non-accreting YSOs \citep{Preibisch2005}. 
Initial explanations for this observation were predominantly based on the notion that the accretion process suppresses, disrupts, or 
obscures the X-ray activity. While \cite{Guedel2007} suggested that the disks may be opaque for parts of 
the X-ray spectrum, \cite{Romanova2004} and \cite{Telleschi2007} proposed that the accretion process may 
change the coronal magnetic field structure or even strip off the coronal magnetic field \citep{Jardine2006},
leading to lower X-ray emission. \cite{Telleschi2007} also discussed the possibility that the accreted material 
may cool the coronal plasma and soften the X-ray emission, making it undetectable for charge-coupled device (CCD) X-ray detectors.

An alternative model introduced by \cite{Drake2009} and based on hydrostatic X-ray photoevaporation calculations 
suggests that the X-rays modulate the accretion flow according to the X-ray heated disk models from \cite{Ercolano2008b, Ercolano2008a, Ercolano2009}. 
The ionizing radiation from the central protostar heats up the gas in the surface layers of the disk. If the 
temperature exceeds the local escape temperature, the hot gas establishes a thermally driven wind and escapes from the disk. 
Far-ultraviolet ($6~\mathrm{eV} \le E < 13.6~\mathrm{eV}$) and soft X-ray ($0.1~\mathrm{keV} < E < 2~\mathrm{keV}$) 
radiation is expected to drive the strongest winds by heating up regions with H column densities on the order of 
$10^{21}~\mathrm{cm}^{-2}$ to $10^{22}~\mathrm{cm}^{-2}$. The mass loss associated with the winds interrupts the supply 
of material into the inner parts of the disk, which is still viscously accreting onto the star. After a few million 
years, the accretion rate has declined to values lower than the mass loss rate in the context of viscous accretion, and a gap forms in the disk.
At this stage, erosion sets in and the accretion rate drops quickly. Stars with higher X-ray luminosities should arrive
earlier at this stage and should therefore display lower accretion rates for a given age. A more detailed outline of the process 
is given in the review of \cite{Ercolano2017}.

In order to test this theory, several studies have been carried out: \cite{Drake2009} found an anticorrelation
between the accretion rates and the mass-normalized X-ray luminosities in a sample of $44$ young stars in the Orion Nebula Cluster (ONC). 
Previously, the \cite{Telleschi2007} study of $39$ sources in the Taurus Molecular Cloud 
using data from the \textit{XMM-Newton Extended Survey} pointed toward a similar relation.
However, due to the small sample sizes, the results of these studies remained tentative. \cite{Bustamante2016} seemingly confirmed the results
with a larger sample of $431$ sources in the ONC.

Due to the importance of the $\dot{M}_\mathrm{acc} - L_\mathrm{X}$ relation, we have performed a new analysis and checked the methods of \cite{Telleschi2007}
and \cite{Bustamante2016} for possible biases. In particular, we have tested the hypothesis that young stars with similar masses and ages show lower accretion rates 
for higher X-ray luminosities, on average, using an unbiased partial regression analysis. Furthermore, we have tested how flaring activity may influence the $\dot{M}_\mathrm{acc} - L_\mathrm{X}$ relation and studied
the relation between the hardness of the X-ray spectrum and the accretion rates.


\section{Sample data and selection }

\subsection{Source data}
With a distance of $\sim 403~\mathrm{pc}$ \citep{Kuhn2019} from our Sun, 
the ONC is one of the nearest regions of ongoing star formation \citep{Bally2008}. The deepest and longest X-ray 
observation of a young stellar cluster in the history of X-ray astronomy is the COUP observation of the ONC with \textit{Chandra}
and the \textit{Advanced CCD Imaging Spectrometer} (ACIS).
An $838~\mathrm{ks}$ 
exposure was made over a period of $13.2$ days in January 2003, producing a data set of $1616$ detected X-ray sources 
\citep{Getman2005b}. The fluxes for $1543$ sources were obtained by integrating thermal plasma spectral fits. 
The total band luminosities for the $73$ remaining, faintest, sources were estimated from the incident fluxes and the median energies. For our analysis, we used the total X-ray luminosity corrected for interstellar absorption (corresponding to the total $0.5 - 8.0\,\mathrm{keV}$ band). 

\cite{Manara2012} determined accretion rates for $730$ young stars with the photometric catalog from the \textit{HST 
Treasury Program} \citep{Robberto2013}. They used the \textit{Wide Field and Planetary Camera 2} (WFPC2) observations carried out between October 2004 
and April 2005 with the F336W (U), F439W (B), F656N (H$\alpha$), and F814W (I) filters. The photometric data were used to construct a 
$U-B$ versus $B-I$ color-color diagram. An isochrone representing the pure photometric emission without 
accretion was constructed using the synthetic BT-SETTL atmospheric model spectra from \cite{Allard2011}, which were calibrated for the 
optical range. It was then assumed that the displacement of the observed sources from this isochrone is due 
to a combination of extinction and accretion. Using the $T_\mathrm{eff}$ estimates from \cite{DaRio2012}, 
both $A_V$ and $L_\mathrm{acc}$ could be uniquely determined. With this procedure, the accretion luminosities 
for $271$ sources were determined (the ``U-excess sample''). These data were further used to obtain a relation between the
accretion luminosities and the H$\alpha$ luminosities, from which the remaining $459$ accretion luminosities 
were derived (the ``H$\alpha$ sample''). This was possible because the emission in the hydrogen lines is linked to the 
accretion process \citep[e.g.,][]{Hartmann2016}.

For our analysis, we adopted the stellar masses determined by \cite{Manara2012} using the evolutionary 
models from \cite{Siess2000}. However, we also repeated the analysis with the models from \cite{Dantona1994} and
\cite{Palla1999} for comparison.

\begin{figure*}
\resizebox{\hsize}{!}
        {
                \includegraphics[width=\hsize]{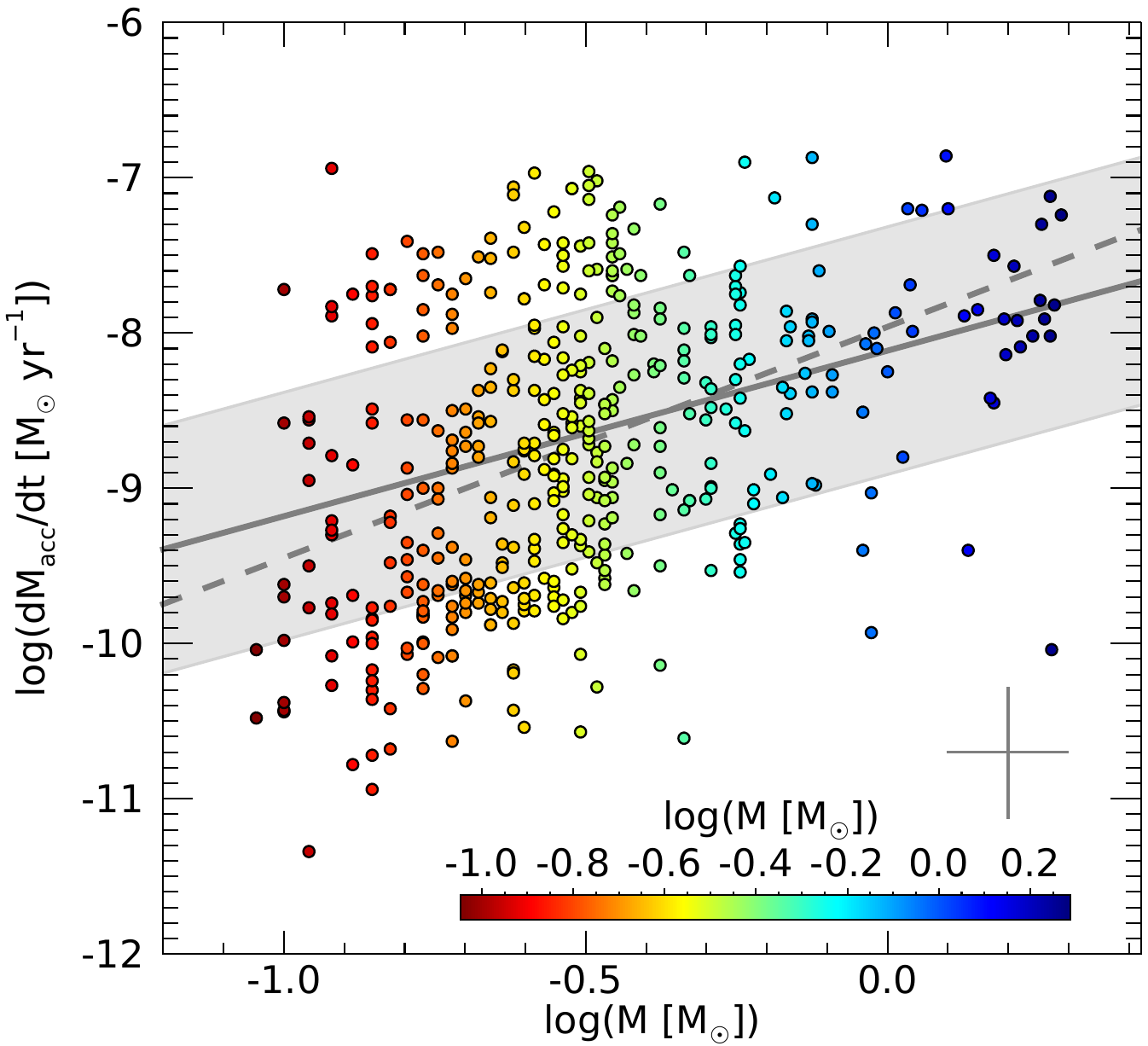}    
                \includegraphics[width=\hsize]{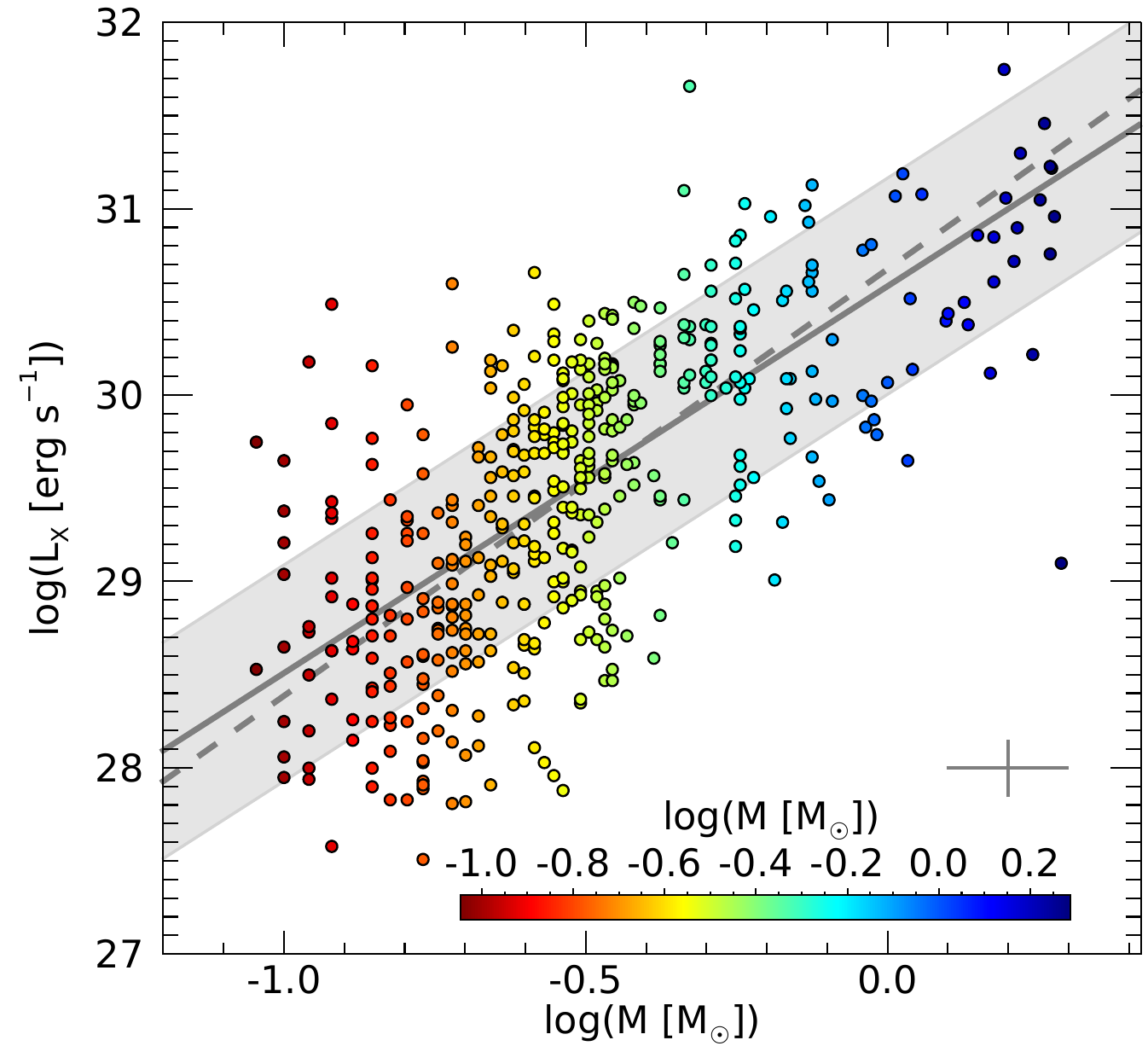}      
        }
        \caption{Correlation of the mass accretion rate and the X-ray luminosity with the stellar mass. Left: Mass accretion rate vs. stellar mass for young stars in the ONC. 
                        The accretion data and the masses are taken from \cite{Manara2012}. The dashed line shows a single 
                        power-law fit for all objects with masses $\leq 2.0~M_\odot$. The solid line indicates the fitting 
                        result for the mass range $0.2 \leq M \leq 2.0$, and the shaded region is its $1$-$\sigma$ deviation 
                        from the data. The cross indicates the typical uncertainty of the values. Right: Same but for the X-ray luminosities taken from \cite{Getman2005}. The fitting parameters are 
                        listed in Table \ref{tab:results}.
        } \label{fig:figmass}
\end{figure*}

\subsection{Crossmatching} \label{sec:sample}

The crossmatching between the two catalogs was performed with the \verb|match_xy| routine of 
the \verb|tara-package|\footnote{\url{http://personal.psu.edu/psb6/TARA/}} as described in \cite{Broos2011b}. 
This routine uses the individual position uncertainties assuming a Gaussian distribution in order to
determine a maximum acceptable distance such that $\sim 99\%$ of the counterparts in the two catalogs are identified as matches. The positional uncertainties for the X-ray sources were taken from the COUP table,
and we assumed a positional uncertainty of $0.1\arcsec$ for the HST sources. The most significant match
of each X-ray source is denoted the ``primary match,'' while less significant matches are labeled ``secondary matches.''
After this classification, the algorithm resolves many-to-one and one-to-many relations. Unambiguous one-to-one
relationships are labeled ``successful primary matches.'' For our sample, this procedure led to $499$ successful primary matches.

\subsection{Gaia distances of the stars in the sample}

From the $499$ primary matches, $483$ sources have parallaxes listed in the Gaia Early Data Release 3 (EDR3) catalog \citep{Gaia2016b, Gaia2020b}.
The mean parallax of the sources amounts to $\left\langle \varpi \right\rangle = 2.52 \pm 0.01~\mathrm{mas}$, corresponding 
to a mean distance of $\approx 397.0~\mathrm{pc}$, which agrees with the distance of $403^{+7}_{-6}~\mathrm{pc}$ determined by 
\cite{Kuhn2019} using the earlier Gaia Data Release 2 (DR2).

We used the Gaia parallax data in order to identify stars that are likely not members of the ONC. In a 
first step, we selected candidates with a $5-\_\varpi$ parallax range that does not contain the mean 
distance. Next, we calculated the individual distances of the candidates. We defined candidates as 
foreground or background stars if their distances do not overlap, within $1\sigma$, with the distance of the cluster 
determined by \cite{Kuhn2019}. Using these criteria, we identified $11$ targets in our sample that are most likely not members of the ONC: 
COUP 180 ($93.5^{+8.8}_{-7.4}~\mathrm{pc}$),
COUP 153 ($114.1^{+0.8}_{-0.8}~\mathrm{pc}$),
COUP 1569 ($140.5^{+1.7}_{-1.6}~\mathrm{pc}$),
COUP 518 ($213.5^{+16.6}_{-14.4}~\mathrm{pc}$),
COUP 958 ($233.7^{+2.3}_{-2.3}~\mathrm{pc}$),
COUP 1516 ($374.5^{+3.4}_{-3.3}~\mathrm{pc}$),
COUP 173 ($493.5^{+24.6}_{-22.4}~\mathrm{pc}$),
COUP 855 ($523.4^{+18.6}_{-17.3}~\mathrm{pc}$),
COUP 1455 ($535.1^{+35.2}_{-31.1}~\mathrm{pc}$),
COUP 847 ($544.9^{+42.9}_{-37.0}~\mathrm{pc}$), and COUP 1585 ($1532.1^{+1997.1}_{-553.7}~\mathrm{pc}$).
These targets were removed from our sample, leaving a total of $488$ sources.

In order to be consistent with the data from \cite{Manara2012}, we scaled the X-ray luminosity to the 
distance of $414~\mathrm{pc}$ \citep{Menten2007}. This does not affect our correlation analysis since the
exact distance does not influence the slope of the relations.

\subsection{Data selection}

Following \cite{Drake2009}, we set an upper mass limit of $M = 2.0\,M_\odot$. This reduced our sample to $453$ sources.
Like\ \cite{Manara2012}, we further restricted the isochronal age interval according to
$5.5 \leq \log(\tau [\mbox{yr}]) \leq 7.3$ in order to reduce the amount of outliers in the Hertzsprung-Russell (HR) diagram in our sample. The final sample contains $446$ sources.

\subsection{Uncertainties of the quantities}

The X-ray luminosities were
derived from fitting spectral models, which are highly nonlinear \citep{Preibisch2005}. Following the
authors, we assumed a typical uncertainty of $0.15~\mathrm{dex}$, based on the typical uncertainties 
derived from the spectral fits.

The accretion rates are subject to several uncertainties since they are calculated from
values that are themselves subject to substantial uncertainties, such as the extinction and the stellar mass.
We assumed typical uncertainties of $0.10~\mathrm{dex}$ for the stellar masses and $0.42~\mathrm{dex}$ for the
accretion rates, as estimated by \cite{Alcala2017} for YSOs in Lupus. The analysis of these authors suggests
that the typical uncertainties are not substantially affected by the choice of the evolutionary models.


\section{Correlation analysis}

\subsection{Relation between accretion rate, X-ray luminosity, and stellar mass}

The left-hand side of Fig. \ref{fig:figmass} shows the (logarithmic) accretion rate as a function of the (logarithmic) stellar mass. In the first step, we followed the guideline of \citet{Isobe1990} and determined the slopes, $\beta$, of the regression line using 
different linear regression methods implemented in the \textit{Interactive Data \mbox{Language}} (IDL) routine \verb|SIXLIN|. We found values of $1.34 \pm 0.14$ (ordinary least squares [OLS]), $2.48 \pm 0.15$ (OLS bisector), $7.39 \pm 0.80$ (orthogonal reduced major axis), $3.27 \pm 0.42$ (reduced major axis), and $4.67 \pm 0.39$ (mean OLS).
Since the slopes differ even within the range of their uncertainties, we conclude that the assumptions on which some of the methods are based are not 
fulfilled by our sample. Therefore, we determined the slopes using the fully Bayesian method from \citet{Kelly2007}, which takes the uncertainties of both variables into account. For this task, the IDL implementation \verb|LINMIX_ERR| was used. The procedure yields a slope of $1.49 \pm 0.16$,
which is consistent with the slope determined with the OLS method. 

The dispersion of the regression line is $0.84~\mathrm{dex}$. This large scatter can be explained by measurement uncertainties and intrinsic variations. The slope agrees, within $1\sigma$, with the value of $1.57 \pm 0.23$ that \cite{Bustamante2016} found for their H$\alpha$-sample,
but it is lower than the value of $1.94 \pm 0.23$ that they found for their U-excess sample. 
The value agrees with the slope of $1.8 \pm 0.2$ that \cite{Alcala2017} determined for Lupus and is at the lower end of the range of the typical findings of $1.5$ to $3.1$ from other studies \citep{Hartmann2016}.

When using the older tracks from \cite{Dantona1994}, the slope becomes $1.77 \pm 0.22$, which is in agreement with the findings of other studies. In particular, this slope agrees with the value of $1.66 \pm 0.07$ that \cite{Ercolano2014} found in their 
total sample, which also contains the accretion rates we used in our analysis. According to the authors, slopes of $1.45-1.70$ are expected in the context of X-ray driven photo-evaporative disk dispersal.

However, newer studies of the $\dot{M}_\mathrm{acc}-M$ relation provide evidence for a break in the distribution 
between $M = 0.1~M_\odot$ and $M = 0.3~M_\odot$. \cite{Manara2017} showed statistically that a double power law describes the $\dot{M}_\mathrm{acc} - M$ relation slightly better than
a single power law for a sample of accreting stars in Chamaeleon  I. The slope is steeper for stars with masses lower than $\sim 0.3~M_\odot$ and flatter for stars with higher 
masses. A similar behavior was found by \cite{Alcala2017} for their sample in Lupus. They observed a steeper slope for stars with masses lower than $0.2~M_\odot$. Possible 
explanations provided by the authors are a faster evolution for objects with lower masses \citep{Manara2012} or two different accretion regimes for different masses \citep{Vorobyov2009}.

\begin{figure*}
\resizebox{\hsize}{!}
        {
                \includegraphics[width=\hsize]{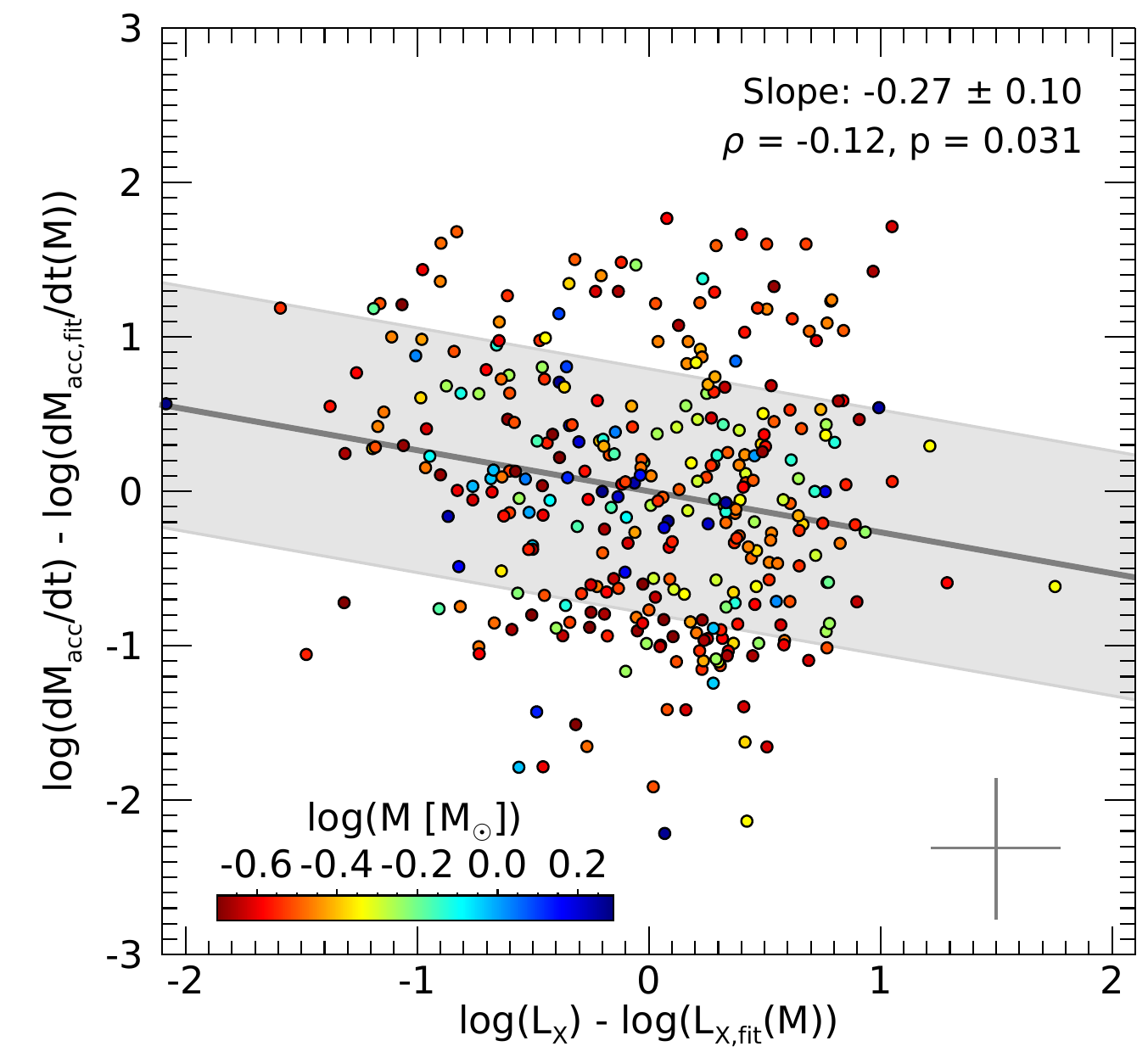}     
                \includegraphics[width=\hsize]{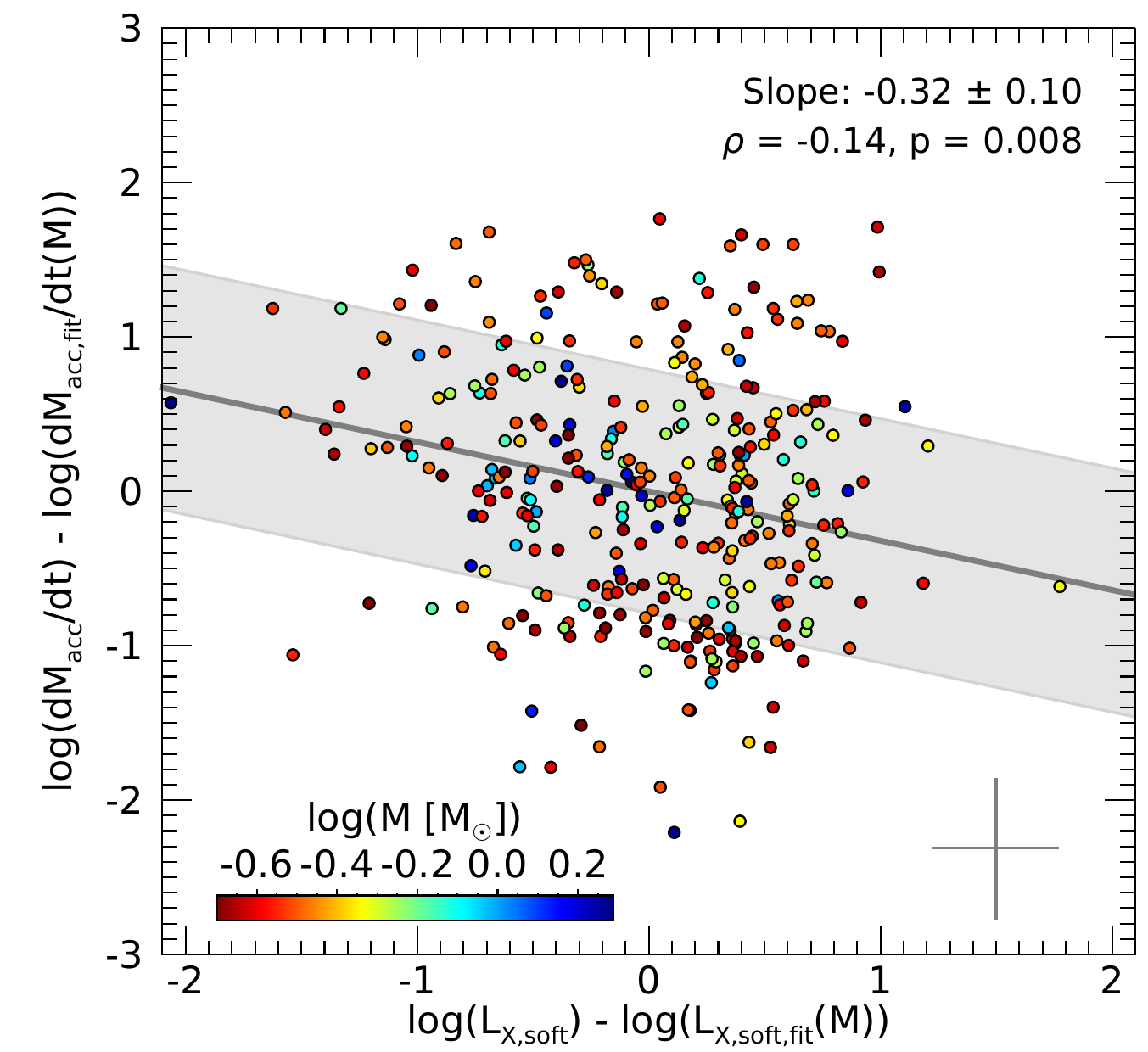}    
        }
        \caption{Correlation of the residual mass accretion rate and the residual X-ray luminosity. Left: Residual mass accretion rate as a function of residual X-ray luminosity. The 
                        regression line was obtained with the Bayesian \mbox{LINMIX\_ERR} method, and the shaded 
                        region shows its $1\sigma$ scatter. The cross indicates the typical uncertainties. 
                        There is a weak anticorrelation. Right: Residual mass accretion rate as a function 
                        of the residual soft band ($0.5 - 2.0~\mathrm{keV}$) X-ray luminosity. The 
                        anticorrelation is stronger and more significant.
        } \label{fig:figmacclx}
\end{figure*}
        
The distribution of our sample suggests a similar behavior: Performing the linear regression using \verb|LINMIX_ERR| only for objects with masses $\ge 0.2~M_\odot$ leads to the relation
\begin{equation}
\log\left(\frac{\dot{M}_\mathrm{acc}}{M_\odot\,\mathrm{yr\,}^{-1}}\right) = (-8.11 \pm 0.10) + (1.07 \pm 0.22) \times \log\left(\frac{M}{M_\odot}\right), \label{eq:eqMacc}
\end{equation}
with a scatter of $0.80$. The slope of $1.07 \pm 0.22$ is flatter compared to the fitting results for the full sample. 
Within $1\sigma$, the slope agrees with the value of $1.37 \pm 0.24$ that \cite{Alcala2017} found for objects with $M \ge 0.2~M_\odot$ in Lupus, but is steeper than
the slope of $0.56$ that \cite{Manara2017} found for $M > 0.3~M_\odot$ in Chamaeleon I.

The right-hand side of Fig. \ref{fig:figmass} shows the X-ray luminosity as a function of 
the stellar mass. The dashed line represents the linear fitting result for the whole sample and has a slope of $2.29 \pm 0.12$, with a spread of $0.61$. Limiting the mass range to values
$\geq 0.2~M_\odot$ leads to a relation with a slightly flatter slope of $2.08 \pm 0.16$ and a spread of $0.58$. The whole relation
reads
\begin{equation}
\log\left(\frac{L_\mathrm{X}}{\mathrm{erg\,s}^{-1}}\right) = (30.58 \pm 0.07) + (2.08 \pm 0.16) \times \log\left(\frac{M}{M_\odot}\right). \label{eq:eqLx}
\end{equation}
The slope is in agreement with the findings from \cite{Bustamante2016} of $1.90 \pm 0.17$ (H$\alpha$-sample), but higher than the 
value of $1.72 \pm 0.17$ (U-excess sample) and the value of $1.69 \pm 0.11$ found by \cite{Telleschi2007} for stars in the Taurus molecular cloud. Furthermore, the slope is larger than the value of 
$1.28 \pm 0.07$ found by \cite{Ercolano2014} for their total sample and the $1.44 \pm 0.10$ obtained by \cite{Preibisch2005} despite the fact that they used the same 
evolutionary tracks. Repeating the analysis with the mass estimates listed in the COUP table, the slope becomes $1.67 \pm 0.21$, in agreement with the slope reported by \cite{Preibisch2005}.

This indicates that the difference of the slopes is not caused by selection effects. In addition to the uncertainties in the $T_\mathrm{eff}$ and $L_\mathrm{bol}$ values used to determine the masses in the HR diagram, there is also a bias that tends to overestimate the masses for brighter sources, resulting in a flattened relation between the masses and the X-ray luminosities. The bolometric luminosities used by \cite{Manara2012} were corrected for 
the accretion luminosities, which are particularly high for heavier stars. Furthermore, the ratio $L_\mathrm{acc} / (L_\mathrm{acc} + L_\mathrm{bol})$ increases
slightly with increasing mass (see Appendix \ref{sec:prop}). Without this correction, the stellar luminosities are overestimated and the positions of the stars in the HR diagram are shifted toward tracks associated with higher masses. 
Since it is known that the X-ray luminosity is positively correlated with the bolometric luminosity \citep[e.g.,][]{Preibisch2005}, the masses of brighter X-ray sources are overestimated when the bolometric luminosity is not corrected for the accretion luminosity. It is possible that the $L_\mathrm{X}-M$ relations found for other star forming regions may suffer the same effect.

Since low mass stars are intrinsically fainter, which gives rise to larger uncertainties, we limited further analysis to objects with masses $\ge 0.2~M_\odot$. Although this restriction reduces the sample size to $332$, it avoids biases arising from the possible bi-model behavior of the $\dot{M}_\mathrm{acc}-M$ relationship. Furthermore, this choice reduces the risk of a possible selection bias: The sample from \cite{Manara2012} is expected to be representative of the ONC stellar population down to $\sim 0.1~M_\odot$. The COUP sample is thought to be complete down to $\sim 0.2~M_\odot$ for lightly absorbed ($A_V < 5~\mathrm{mag}$) stars \citep{Preibisch2005}. We are therefore confident that our analysis will not be affected by a selection bias.

Sources from the U-excess sample make up $39\%$ of our final sample, with sources from the H$\alpha$ sample making up the remaining $61\%$. In order to quantitatively test the correlation between the parameters, we used the Spearman correlation coefficient, $\rho$, implemented in the IDL routine \verb|R_CORRELATE|.
It is $0.62$ for the $L_\mathrm{X}-M$ and $0.32$ for the $\dot{M}_\mathrm{acc}-M$ relation. The two-sided significance of its 
deviation from zero (the ``p-value'') is $< 0.001$ for both cases, that is, the probability that the results are not due to a correlation is lower than $0.1 \%$. The linear correlation coefficient, $r$, determined with \verb|LINMIX_ERR| is $0.62 \pm 0.04$ and $0.32 \pm 0.06$, respectively.

It is worth noting that the slope of the $L_\mathrm{X}-M$ relation is always higher than the slope of the $\dot{M}_\mathrm{acc}-M$ relation in our sample, given the mass and isochronal age constraints  of $0.2~M_\odot \le M \le 2.0~M_\odot$ and $5.5 \le \log(\tau / \mathrm{yr}) \le 7.3$, not only for the evolutionary tracks from \cite{Siess2000} but also for the models from \cite{Dantona1994} and \cite{Palla1999}.
Depending on the chosen model, the slope of the $L_\mathrm{X}-M$ relation is larger than the slope of the $\dot{M}_\mathrm{acc}-M$ relation by factors of $1.6-2.0$. Thus, a comparison of the connection between the slopes 
of the $L_\mathrm{X}-M$ and the $\dot{M}_\mathrm{acc}-M$ relations with current model predictions remains inconclusive with the current data set. A new theoretical investigation of the dependence of X-ray photoevaporation mass loss rates on stellar mass is in progress (Picogna et al., in preparation) and will address this point in more detail. 

\subsection{Relation between X-ray activity and accretion rate} \label{sec:macclx}

The main goal of this work is to check the hypothesis that, in a sample with similar masses and ages, stars 
with higher X-ray luminosities show lower accretion rates on average. Since both $L_{\mathrm{X}}$ and $\dot{M}_\mathrm{acc}$ display a correlation with the stellar mass, one has to take their relation into account when analyzing the relation between $L_{\mathrm{X}}$ and $\dot{M}_\mathrm{acc}$. 

For the sake of readability, the following short notations are introduced: $X$  stands for the logarithm of the X-ray luminosity, $Y$ for the 
logarithm of the accretion rate, and $m$ for the logarithm of the stellar mass.
In order to illustrate the problem concerning the common mass dependence, a simple model consisting of two relations, $X_0$ and $Y_0$  -- both of which are linearly 
dependent on a third variable, $m$ -- is regarded. Furthermore, scatter terms ($s_x$ and $s_y$) are added. The scatter may be due to 
measurement uncertainties and/or the intrinsic variation of $X_0$ and $Y_0$ for identical $m$. The model is described by the following relations:
\begin{eqnarray}
X_0 & = & a + b \cdot m \label{eq:eqX0} ,\\
Y_0 & = & c + d \cdot m \label{eq:eqY0} ,\\
X & = & X_0 + s_x = a + b \cdot m + s_x \label{eq:eqX} ,\\
Y & = & Y_0 + s_y = c + d \cdot m + s_y \label{eq:eqY}
.\end{eqnarray}
Here, $X_0$ and $Y_0$ are the relations one would ideally obtain from a linear regression of the $X-m$ relation and the $Y-m$ relation, respectively. 

Since both $X$ and $Y$ depend on $m$, the $X-Y$ relation will display a correlation, even if there is no intrinsic connection between 
the two quantities. The slope of this relation is $d / b$, which can easily be shown. From Eqs. \ref{eq:eqX} and \ref{eq:eqY}, it follows that
\begin{equation*}
Y(X) = c + s_y + \frac{d}{b} \cdot (X - a - s_x),
\end{equation*}
which is a line with a slope of $d / b$. From this exercise, we can conclude that a correlation analysis will suffer spurious correlations if the 
parameters under consideration depend on a common additional variable, referred to as the ``common response variable'' in the literature \citep[for instance]{Shirley2010}.

To overcome this problem, one usually resorts to a partial or semi-partial regression analysis \citep{Wall2003, Dey2019, Drake2009}, which basically consists of two steps: 
First, the residua of one or more variables are calculated by taking the difference between the observed values and the linear regression solutions
regarding the variables and the common response variable. Second, the correlation between the residua (partial correlation) or between the 
residua and the variables (semi-partial correlation) is investigated.

In our model, we know the linear regression solutions regarding $m$ and can immediately state that the residua are $X - X_0 = s_x$ and $Y - Y_0 = s_y$.
Unlike the previous case, there is no shared variable in this approach.

\begin{figure*}
\resizebox{\hsize}{!}
        {
                \includegraphics[width=\hsize]{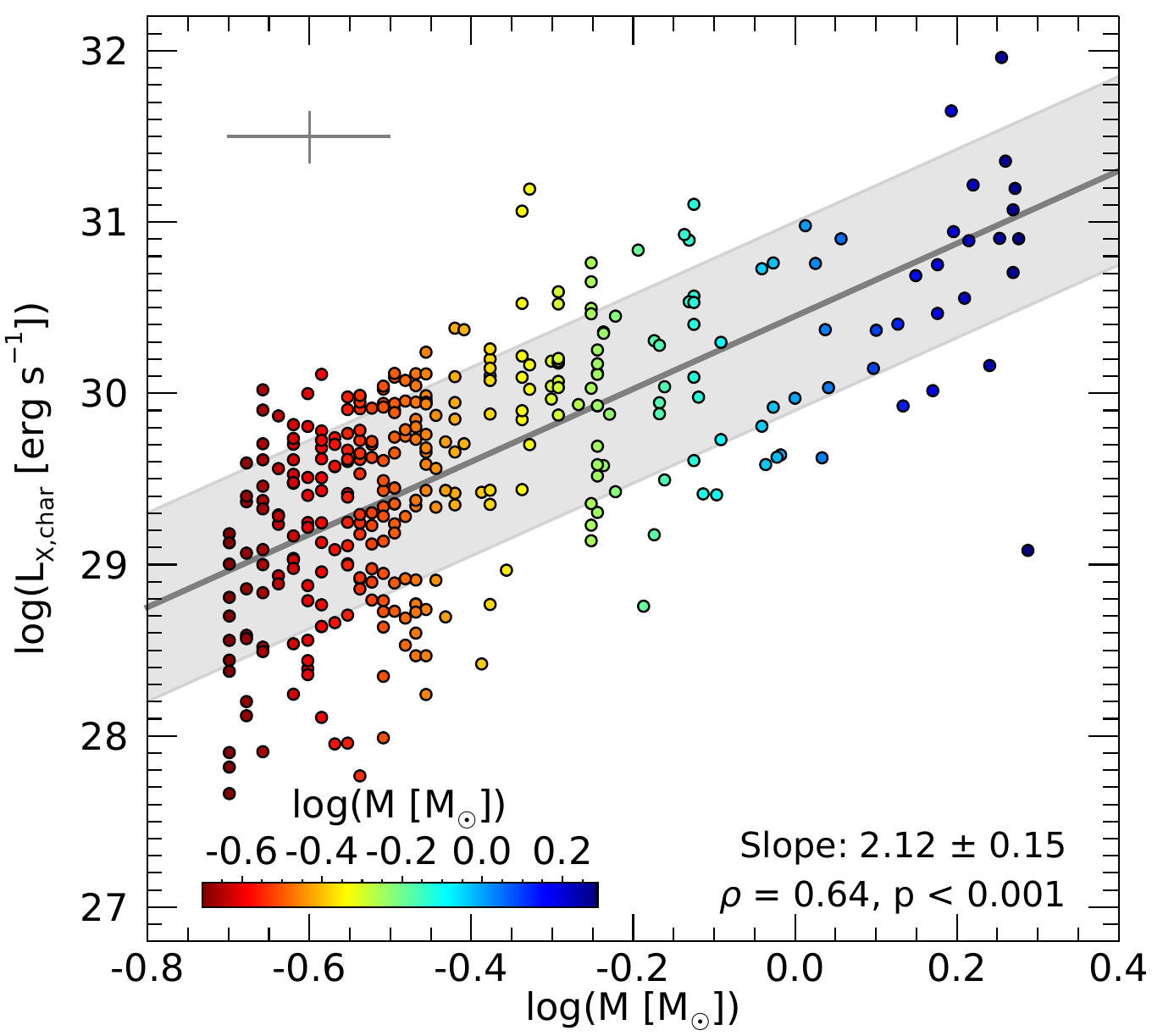}   
                \includegraphics[width=\hsize]{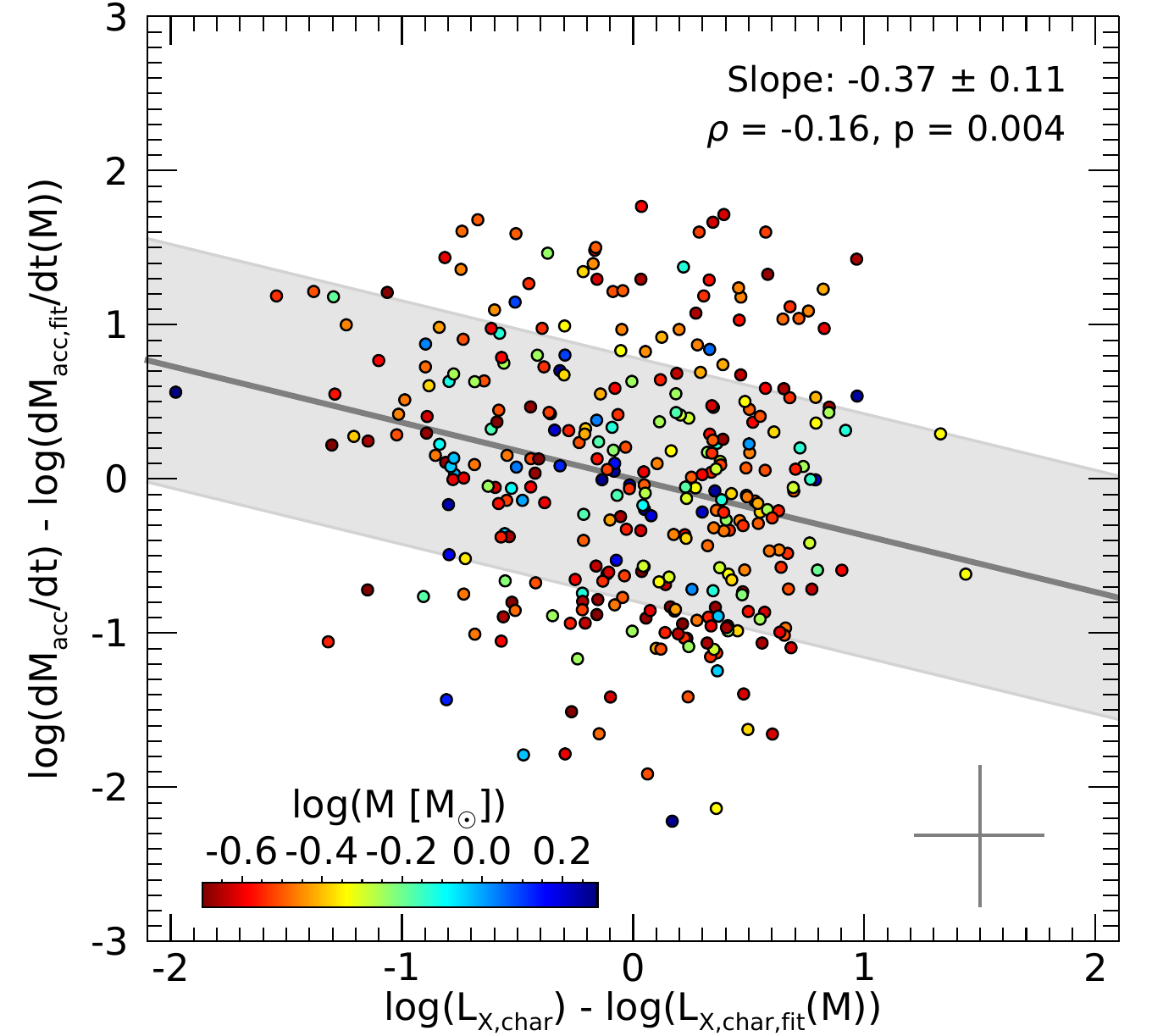}        
        }
        \caption{
				Correlations using the ``characteristic'' X-ray luminosity $L_{\mathrm{X,\,char}}$. Left: $L_{\mathrm{X,\,char}}$ vs. $M$. The cross indicates the typical uncertainties. The $L_{\mathrm{X,\,char}} - M$ relation does not differ significantly from the $L_{\mathrm{X}}-M$ relation (see Fig. \ref{fig:figmass}).
                        Right: Same plot as in Fig. \ref{fig:figmacclx}, but corrected for flaring activity, as described in Sect. \ref{sec:flaring}. The anticorrelation is stronger and more significant.
        } \label{fig:figflare}
\end{figure*}
\begin{figure*}
\resizebox{\hsize}{!}
        {
                \includegraphics[width=\hsize]{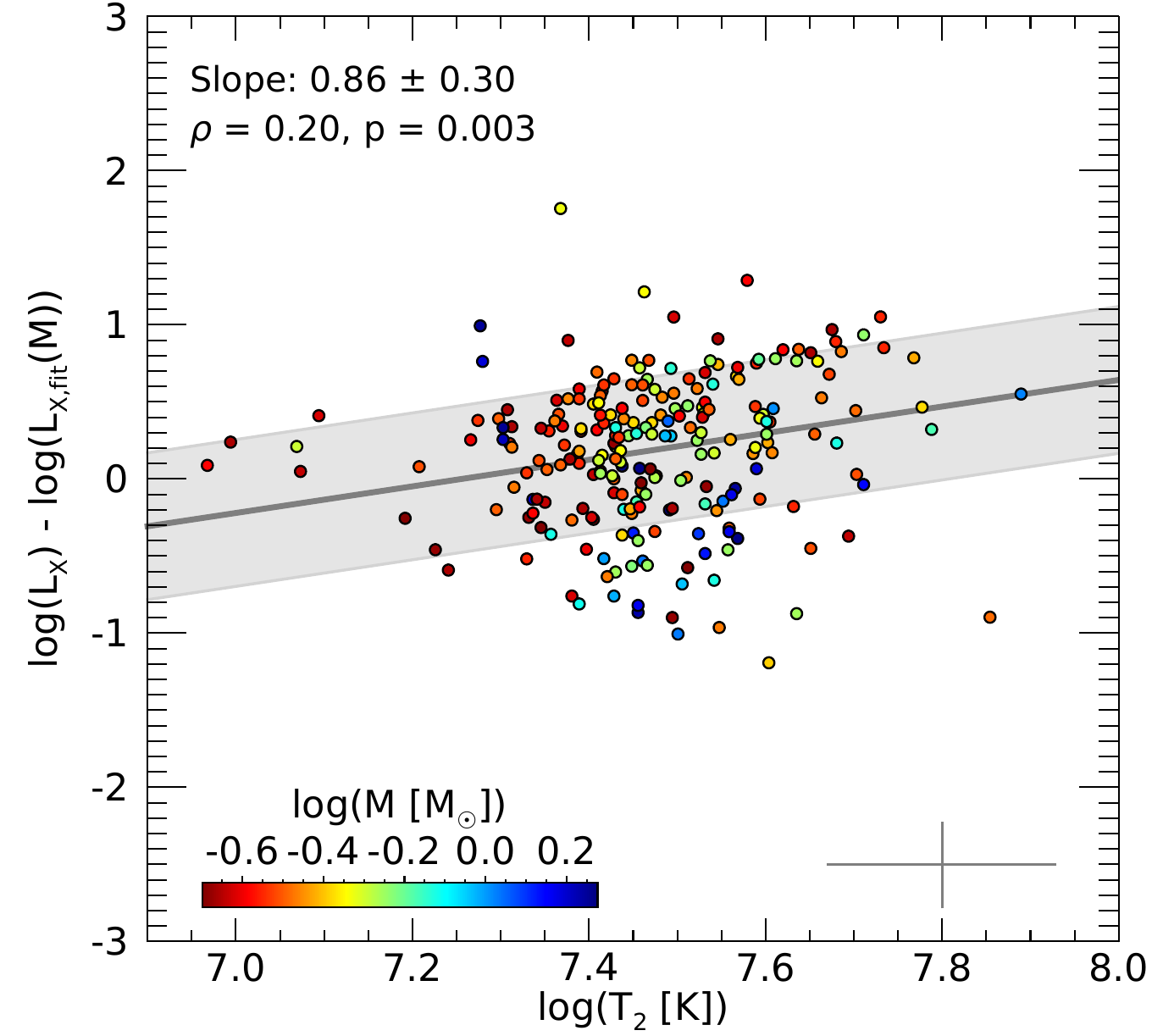} 
                \includegraphics[width=\hsize]{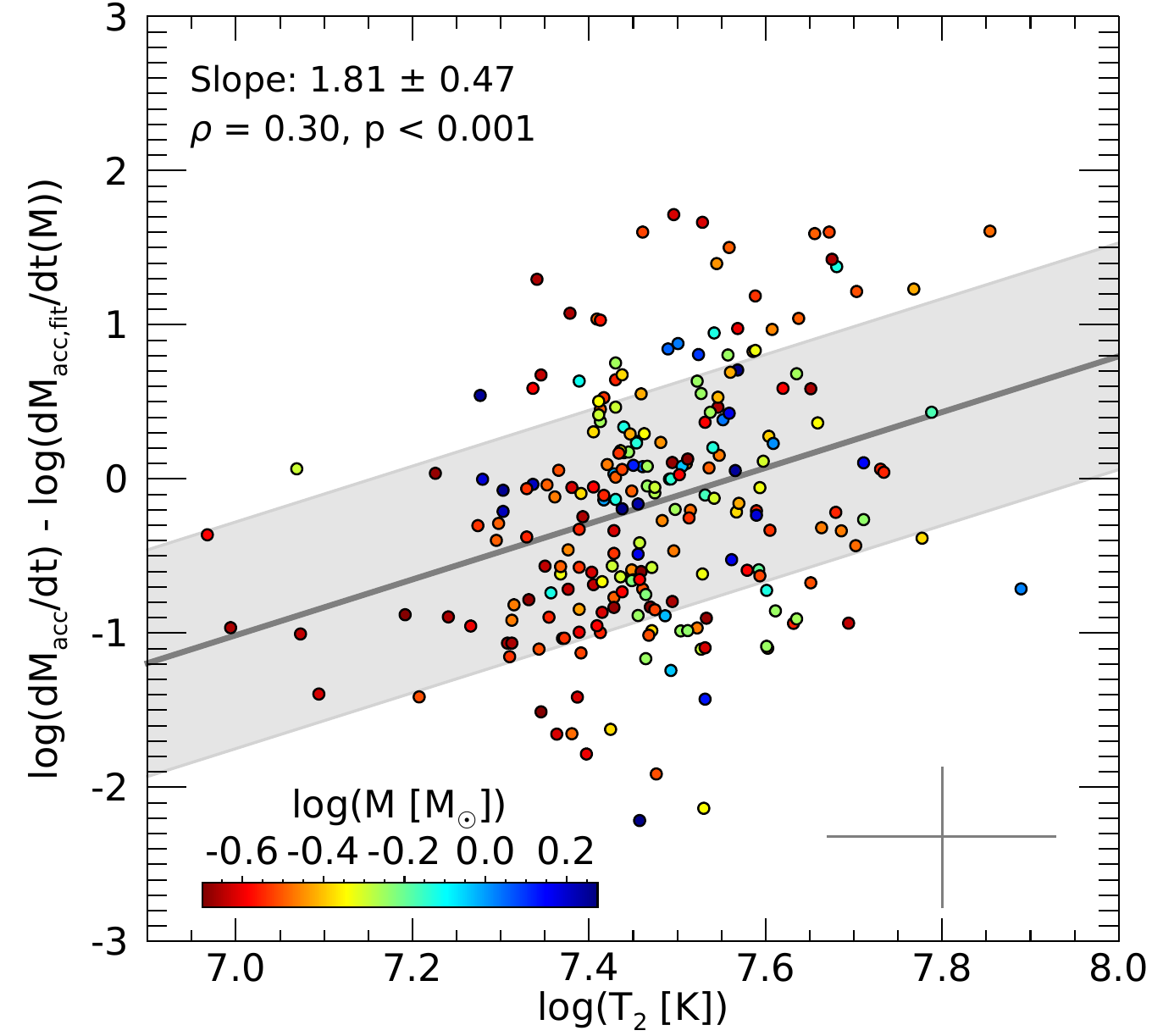}
        }
        \caption{Residual X-ray luminosity (left) and residual mass accretion rate (right) as a function of the higher 
                        plasma temperature, $T_2$, gained in the two-component spectral fitting of \cite{Getman2005}. The cross indicates the typical uncertainties. A significant positive correlation is present in both cases.
        } \label{fig:figplasma}
\end{figure*}

A different method was used by \cite{Telleschi2007} and \cite{Bustamante2016} in their analysis of the relation between $L_\mathrm{X}$ and $\dot{M}_\mathrm{acc}$. Applied to our model, the method works as follows:
The first step is similar to the partial regression analysis; a linear regression on both $X$ and $Y$ with respect to $m$ is performed, resulting in Eqs. \ref{eq:eqX0} and \ref{eq:eqY0}.
In the next step, the theoretical $X$ as a function of $Y$ if both relations only depend on $m$ is constructed. From Eqs. \ref{eq:eqX0} and \ref{eq:eqY0} it follows that
\begin{equation}
X_0(Y_0) = a - \frac{b \cdot c}{d} + \frac{b}{d} \cdot Y_0. \label{eq:XvY}
\end{equation}
Next, the ``residual X'' is calculated as $X - X_0(Y)$. We note that $Y$, not $Y_0$, is inserted into Eq. \ref{eq:XvY}. Finally, a linear regression analysis is performed with the $Y$ - $[X - X_0(Y)]$ relation.
\cite{Telleschi2007} expected the values to scatter around a constant if $[X - X_0(Y)]$ is determined solely by the $X$ - $m$ and $Y$ - $m$ relations. However, this is not the case: From Eqs. \ref{eq:eqX0} to \ref{eq:XvY} one obtains 
\begin{equation}
X - X_0(Y) = s_x - \frac{b}{d} \cdot s_y. \label{eq:xcal}
\end{equation}
Since $Y = Y_0 + s_y$ depends on $s_y$ as well, this method introduces a spurious correlation. Solving Eq. \ref{eq:xcal} for $s_y$ and inserting it into Eq. \ref{eq:eqY} provides
\begin{equation*}
Y[X - X_0(Y)] = c + d \cdot m - \frac{d}{b} \cdot ([X - X_0(Y)] - s_x),
\end{equation*}
implying that the relation will scatter around a line with a slope of $-d/b$ and not a constant. Since both $b$ and $d$ are positive in our case, this approach will inevitably 
introduce a bias toward an anticorrelation, even if there is no intrinsic connection between $X$ and $Y$. In Appendix \ref{a1}, we demonstrate this with a synthetic data set.
The approach would work if $s_y$ were negligible (in that case, the method is essentially equal to the semi-partial regression approach), but this is not the case: The scatter
$s_y$ represents the scatter of the $\dot{M}_\mathrm{acc}-M$ relation, which is particularly large, as shown in the previous section.

In order to avoid this kind of bias, we instead used the partial regression approach with the residual parameters 
$L_{\mathrm{X}} / L_{\mathrm{X}}(M)$ and $\dot{M}_\mathrm{acc} / \dot{M}_\mathrm{acc}(M)$, where $L_{\mathrm{X}}(M)$ and $\dot{M}_\mathrm{acc}(M)$ are given by Eqs. \ref{eq:eqLx} and \ref{eq:eqMacc}. A scatter plot of the values is shown in the left panel of Fig. \ref{fig:figmacclx}. We propagated the typical uncertainties to the residual values and performed the linear regression with the \verb|LINMIX_ERR| routine. We found the following relation:
\begin{equation}
\log\left(\frac{\dot{M}_\mathrm{acc}}{\dot{M}_\mathrm{acc}(M)}\right) = 0.00 \pm 0.04 + (-0.27 \pm 0.10) \times \log\left(\frac{L_{\mathrm{X}}}{L_{\mathrm{X}}(M)}\right).
\end{equation}
Although the dispersion is large with $0.79~\mathrm{dex}$, a weak anticorrelation could be found (Spearman's $\rho$ is $-0.12$ with $p = 0.031$ and a linear correlation coefficient of $r = -0.21 \pm 0.07$), indicating  that, for similar masses, young stars with higher X-ray luminosities show lower accretion rates on average.

Furthermore, the choice of the evolutionary models used for the mass and isochronal age determination
influences the result. When using the models from \cite{Dantona1994}, the slope is $-0.21 \pm 0.12$ ($\rho = -0.11$, $p = 0.082$, and $r = -0.17 \pm, 0.09$), while the models from \cite{Palla1999} lead to a slope of $-0.34 \pm 0.10$ ($\rho = -0.19$, $p = 0.001$, and $r = -0.28 \pm 0.08$).

In Fig. \ref{fig:figmacclx}, the values are color coded according to the masses. A particular trend for different mass regimes is not visible.

The studies carried out by \cite{Telleschi2007} for the Taurus Molecular Cloud and by \cite{Bustamante2016} for the ONC also found 
an anticorrelation between accretion rates and X-ray activity. However, their results suggest a considerably stronger anticorrelation. 
As mentioned above, this is due to a bias introduced in the analysis method.
Indeed, when using the partial regression method, the U-excess sample of \cite{Bustamante2016} displays no correlation 
at all, and the H$\alpha$-sample even shows a slight positive correlation. Combining the two samples and applying the same restrictions as described 
in Sect. \ref{sec:sample} leads to the same weak anticorrelation as found in our analysis if we use the same evolutionary tracks from \cite{Dantona1994}, as \cite{Bustamante2016} did.

\subsection{Influence of the X-ray energy}

Simulations suggest that X-ray photons in the soft band with $E \lesssim 2~\mathrm{keV}$ are particularly efficient in driving disk winds \citep{Ercolano2017}. Therefore, one would expect a stronger anticorrelation between the soft band ($0.5 - 2.0~\mathrm{keV}$) X-ray luminosities, $L_\mathrm{X,\,soft}$, and the accretion rates.
The $L_\mathrm{X,\,soft} - M$ relation, as determined by the \verb|LINMIX_ERR| method, has a flatter slope of $2.00 \pm 0.16$ and a spread of $0.58$.
The residual $\dot{M}_\mathrm{acc} / \dot{M}_\mathrm{acc}(M) - L_\mathrm{X,\,soft} / L_\mathrm{X,\,soft}(M)$ relation shown on the right-hand side of Fig. \ref{fig:figmacclx} displays a slope of $-0.32 \pm 0.10$. With $\rho = -0.14$ ($p = 0.008$) and $r = -0.25 \pm 0.07$, 
the anticorrelation is more significant. The $p$ value is also below $5\%$  for the models from \cite{Dantona1994} ($p = 0.033$) and \cite{Palla1999} ($p < 0.001$). The linear correlation coefficients are $-0.21 \pm 0.09$ and $-0.31 \pm 0.08$, respectively.

\begin{figure*}
\resizebox{\hsize}{!}
        {
                \includegraphics[width=\hsize]{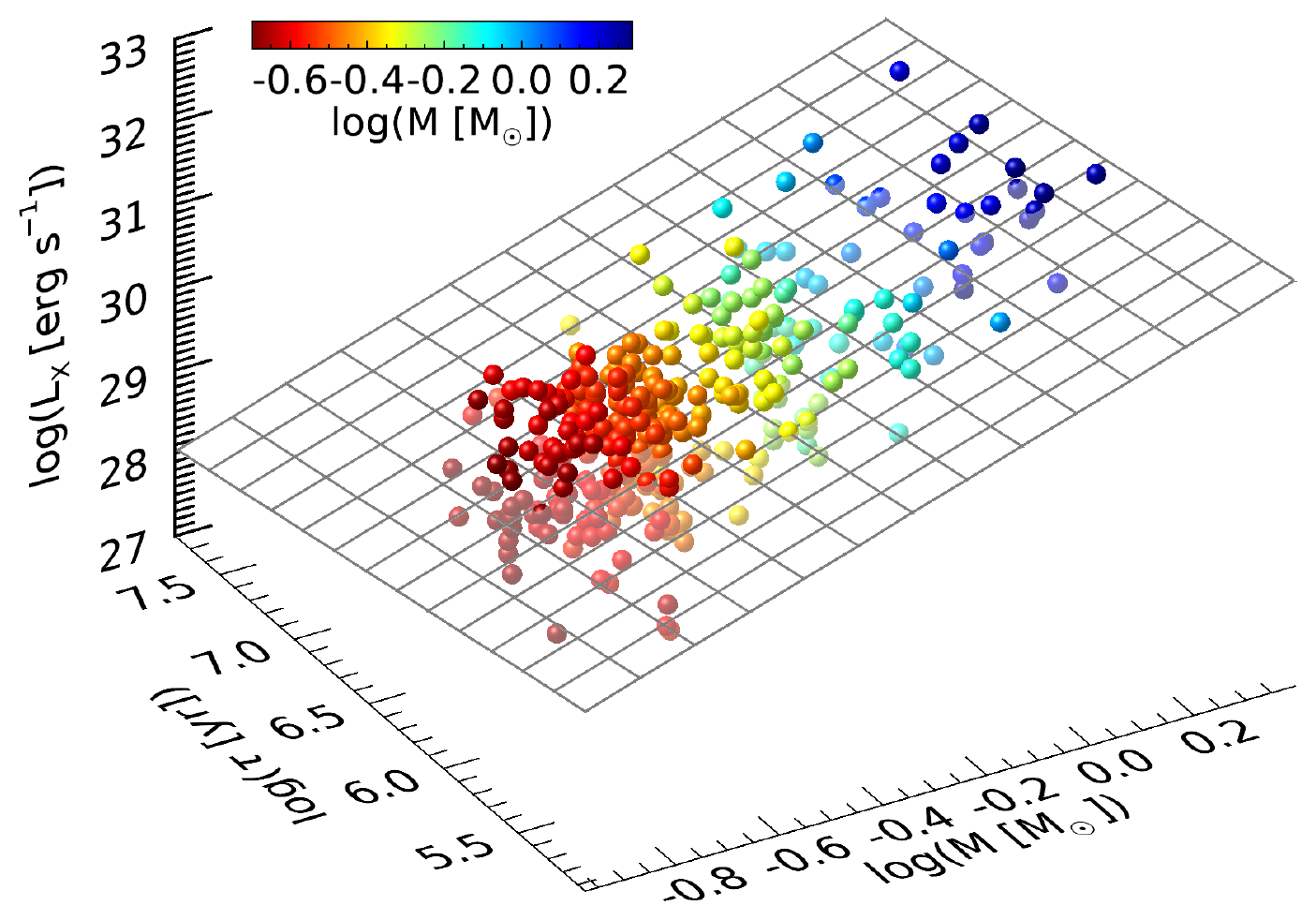}  
                \includegraphics[width=\hsize]{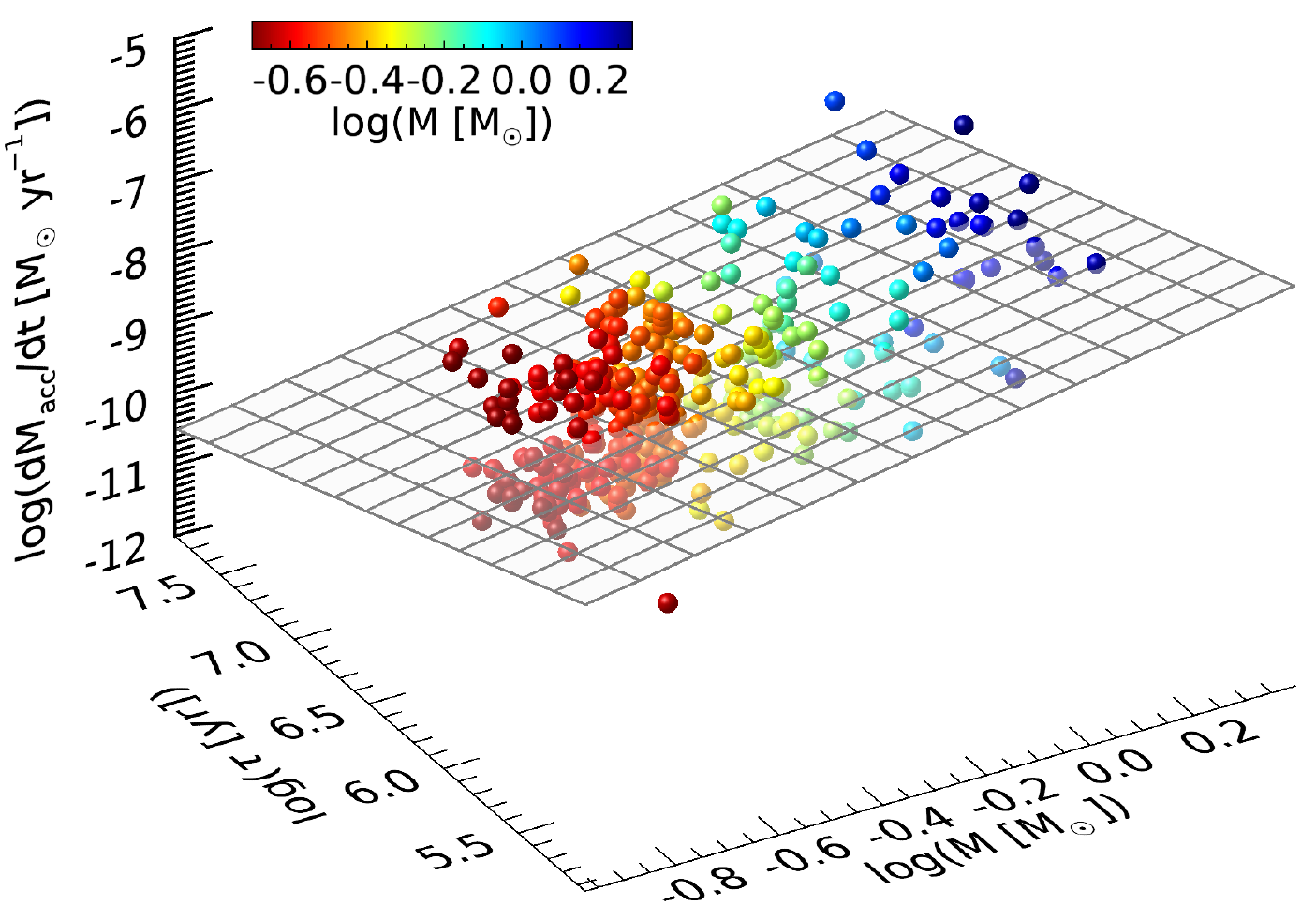}  
        }
        \caption{3D plots of $L_\mathrm{X}$ (left) and $\dot{M}_\mathrm{acc}$ (right) with respect to both $M$ and $\tau$. The plots also show the best-fit surface described by the equation
                $\log Y = A + B \times \log M + C \times \log \tau$, where $Y$ stands 
                for both $L_\mathrm{X}$ and $\dot{M}_\mathrm{acc}$. The coefficients of the fits are listed 
                in Table \ref{tab:tabparams}. The plot points are color coded according to the masses.          
        } \label{fig:fig2d}
\end{figure*}

\subsection{Influence of flaring activity} \label{sec:flaring}

In order to check whether flaring activity has an influence on the regression results, we tried to correct the X-ray luminosities from
flaring influences. \cite{Wolk2005} identified periods of flaring from the light curves of the COUP sources and defined ``characteristic count rates,'' which represent the mean quiescent level of X-ray emission outside the time periods with significant
flares. The ratio between this characteristic count rate and the mean count rate yields a correction
factor, which can be used to determine a ``characteristic'' X-ray luminosity, $L_{\mathrm{X,\,char}}$.

The left-hand side of Fig. \ref{fig:figflare} shows the relation between $L_{\mathrm{X,\,char}}$ and the stellar mass. It displays
a significant positive correlation with $\rho = 0.64$ ($p < 0.001$) and $r = 0.65 \pm 0.04$. The regression parameters of the $L_{\mathrm{X,\,char}}-M$ relation differ only slightly from the respective values of the $L_{\mathrm{X}}-M$ relation and match within their $1$-$\sigma$ uncertainties. With
a spread of $0.55$, the scatter is slightly smaller.
Performing a similar partial regression analysis as the one in Sect. \ref{sec:macclx}, but with the new relations, leads to 
\begin{equation}
\log\left(\frac{\dot{M}_\mathrm{acc}}{\dot{M}_\mathrm{acc}(M)}\right) = 0.00 \pm 0.04 + (-0.37 \pm 0.11) \times \log\left(\frac{L_{\mathrm{X,\,char}}}{L_{\mathrm{X,\,char}}(M)}\right),
\end{equation}
with $\rho = -0.16$ ($p = 0.004$) and $r = -0.26 \pm 0.07$. A scatter plot of the values is shown on the right-hand side of Fig. \ref{fig:figflare}. The anticorrelation is stronger 
and more significant compared to the $L_\mathrm{X}-\dot{M}_\mathrm{acc}$ relation. With a spread of $0.79$, the scatter is similar to the
corresponding relation without the flaring correction.
Using the models from \cite{Dantona1994} for the mass and isochronal age estimates, the slope becomes $-0.32 \pm 0.13$ ($\rho = -0.15$, $p = 0.017$, and $ = -0.23 \pm 0.09$), and the models from \cite{Palla1999} lead to a slope of $-0.44 \pm 0.11$ ($\rho = -0.23$, $p < 0.001$, and $-33 \pm 0.07$).
The results suggest that strong flaring may partly disturb the anticorrelation between X-ray luminosity and accretion. The 3D magnetohydrodynamic simulations from \cite{Colombo2019} suggest that 
flaring activity is indeed able to promote or slow down accretion processes. From this scenario, it is expected that including flaring in the analysis will lead to an 
increased scatter, which tends to be higher for larger X-ray luminosities. This results in a flatter slope due to regression dilution.

\begin{table*}
\caption{\label{tab:tabparams}Coefficients of the equation $\log Y = A + B \times \log M + C \times \log \tau$ obtained by fitting $L_\mathrm{X}$, $\dot{M}_\mathrm{acc}$, and $L_\mathrm{acc}$ as a function of both $M$ and $\tau$ using a standard Levenberg-Marquardt regression.}
\centering
\begin{tabular}{lrrrr}
\hline\hline\\[-2.3ex]
Relation\tablefootmark{(1)} & \multicolumn{1}{c}{$A$\tablefootmark{(2)}} & \multicolumn{1}{c}{$B$\tablefootmark{(2)}} & \multicolumn{1}{c}{$C$\tablefootmark{(2)}} & \multicolumn{1}{c}{$\backslash$\tablefootmark{(3)}}  \\
\hline\\[-2.3ex]
$L_\mathrm{X}$ vs. $M$ and $\tau$ & $33.15 \pm 1.30$ & $1.99 \pm 0.23$ & $-0.41 \pm 0.20$ & $0.57$ \\
$\dot{M}_\mathrm{acc}$ vs. $M$ and $\tau$ & $-2.20 \pm 1.30$ & $1.17 \pm 0.23$ & $-0.92 \pm 0.20$ & $0.76$ \\
$L_\mathrm{acc}$ vs. $M$ and $\tau$ & $2.32 \pm 1.31$ & $1.84 \pm 0.23$ & $-0.52 \pm 0.20$ & $0.75$ \\
\hline
\end{tabular}
\tablefoot{
\tablefoottext{1}{The relations correspond to the logarithmic equation of the parameters.}
\tablefoottext{2}{Errors are the standard deviations.}
\tablefoottext{3}{Standard deviation of the residuum (scatter).}
}
\end{table*}

\begin{table*}
\caption{\label{tab:results}Summary of the correlation analysis results. The relations correspond to the logarithmic equation of the parameters and have the form $\log y = a + b \times \log x$. The parameters were determined with the LINMIX\_ERR method.}
\centering
\begin{tabular}{lcrrrrrr}
\hline\hline\\[-2.3ex]
Parameters & \multicolumn{1}{c}{$N$\tablefootmark{(1)}} & \multicolumn{1}{c}{$a$\tablefootmark{(2)}} & 
\multicolumn{1}{c}{$b$\tablefootmark{(2)}} & \multicolumn{1}{c}{$\rho$\tablefootmark{(3)}} & \multicolumn{1}{c}{$p$\tablefootmark{(4)}} & \multicolumn{1}{c}{$r$\tablefootmark{(5)}} & \multicolumn{1}{c}{$\backslash$\tablefootmark{(6)}} \\
\hline\\[-2.3ex]
$L_\mathrm{X}$ vs. $M$ & $332$ & $30.58 \pm 0.07$ & $2.08 \pm 0.16$ & $0.62$ & $< 0.001$ & $0.62 \pm 0.04$ & $0.58$ \\
$L_\mathrm{X,\,soft}$ vs. $M$ & $332$ & $30.39 \pm 0.07$ & $2.00 \pm 0.16$ & $0.60$ & $< 0.001$ & $0.62 \pm 0.04$ & $0.58$ \\
$L_{\mathrm{X,\,char}}$ vs. $M$ & $332$ & $30.45 \pm 0.07$ & $2.12 \pm 0.15$ & $0.64$ & $< 0.001$ & $0.65 \pm 0.04$ & $0.55$ \\
$\dot{M}_\mathrm{acc}$ vs. $M$ & $332$ & $-8.11 \pm 0.10$ & $1.07 \pm 0.22$ & $0.32$ & $< 0.001$ & $0.32 \pm 0.06$ & $0.80$ \\
\hline\\[-2.3ex]
$\dot{M}_\mathrm{acc} / \dot{M}_\mathrm{acc}(M)$ vs. $L_{\mathrm{X}} / L_{\mathrm{X}}(M)$ & $332$ & $0.00 \pm 0.04$ & $-0.27 \pm 0.10$ & $-0.12$ & $0.031$ & $-0.21 \pm 0.07$ & $0.79$ \\
$\dot{M}_\mathrm{acc} / \dot{M}_\mathrm{acc}(M)$ vs. $L_{\mathrm{X,\,soft}} / L_{\mathrm{X,\,soft}}(M)$ & $332$ & $0.00 \pm 0.04$ & $-0.32 \pm 0.10$ & $-0.14$ & $0.008$ & $-0.25 \pm 0.07$ & $0.79$ \\
$\dot{M}_\mathrm{acc} / \dot{M}_\mathrm{acc}(M)$ vs. $L_{\mathrm{X,\,char}} / L_{\mathrm{X,\,char}}(M)$ & $332$ & $0.00 \pm 0.04$ & $-0.37 \pm 0.11$ & $-0.16$ & $0.004$ & $-0.26 \pm 0.07$ & $0.79$ \\
$\dot{M}_\mathrm{acc} / \dot{M}_\mathrm{acc}(M,\tau)$ vs. $L_{\mathrm{X}} / L_{\mathrm{X}}(M,\tau)$ & $332$ & $0.00 \pm 0.04$ & $-0.27 \pm 0.07$ & $-0.19$ & $< 0.001$ & $-0.20 \pm 0.05$ & $0.74$  \\
$L_\mathrm{acc} / L_\mathrm{acc}(M,\tau)$ vs. $L_{\mathrm{X}} / L_{\mathrm{X}}(M,\tau)$ & $332$ & $0.00 \pm 0.04$ & $-0.26 \pm 0.07$ & $-0.18$ & $< 0.001$ & $-0.20 \pm 0.05$ & $0.73$ \\
\hline\\[-2.3ex]
$L_{\mathrm{X}} / L_{\mathrm{X}}(M)$ vs. $T_2$ & $229$ & $-6.26 \pm 2.27$ & $0.86 \pm 0.30$ & $0.20$ & $0.003$ & $0.27 \pm 0.10$ & $0.48$ \\
$\dot{M}_\mathrm{acc} / \dot{M}_\mathrm{acc}(M)$ vs. $T_2$ & $229$ & $-13.68 \pm 3.51$ & $1.81 \pm 0.47$ & $0.30$ & $< 0.001$ & $0.36 \pm 0.09$ & $0.74$ \\
\hline
\end{tabular}
\tablefoot{
\tablefoottext{1}{Sample size}
\tablefoottext{2}{$a$ is the intercept and $b$ the slope of the linear regression line. Errors for the least squares analysis are the standard deviations.}
\tablefoottext{3}{Spearman's $\rho$ rank correlation coefficient.}
\tablefoottext{4}{$p$-value: two-sided significance of $\rho$'s deviation from zero.}
\tablefoottext{5}{Linear correlation coefficient determined with LINMIX\_ERR.}
\tablefoottext{6}{Standard deviation of the residuum (scatter).}
}
\end{table*}

\begin{figure}
\resizebox{\hsize}{!}
        {
                \includegraphics[width=\hsize]{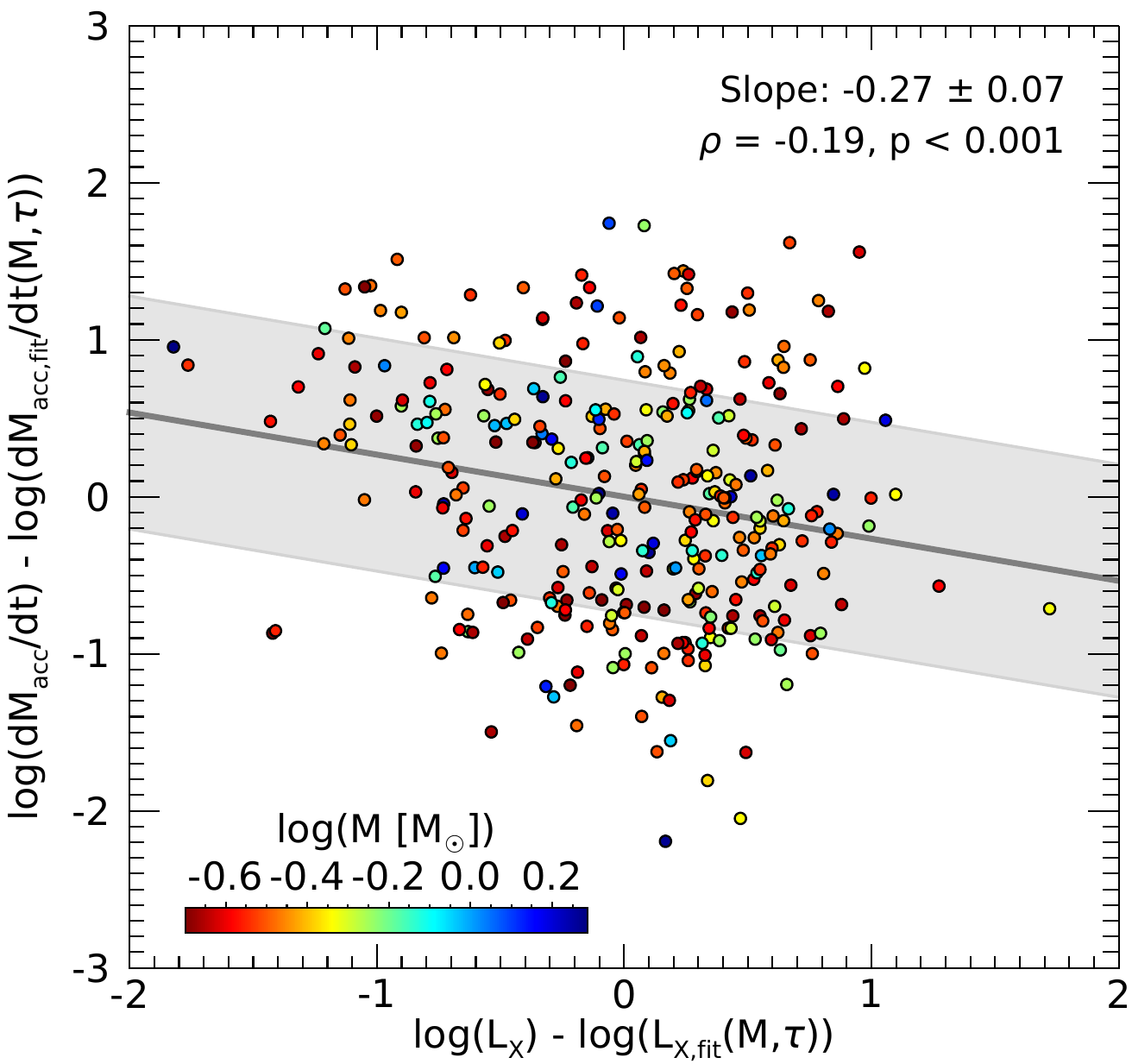}
        }
        \caption{Residual accretion rates vs. residual X-ray luminosities for the ONC taking both the mass and the isochronal age dependences into account.
                                The anticorrelation is stronger and more significant.} \label{fig:figrestau}
\end{figure} 

   \subsection{Influence of stellar age}

It is known that the X-ray luminosity decreases with isochronal age for similar masses \citep[e.g.,][]{Preibisch2005b, Telleschi2007}. A similar behavior can
be observed for the accretion rates \citep[e.g.,][]{Manara2012}. We tried to remove the influence of the common isochronal age dependence from the $M_\mathrm{acc} - L_\mathrm{X}$ relation.
For this purpose, we regarded both the mass and the isochronal age simultaneously by fitting planes to the quantities. Uncertainties were not taken into account during this procedure. Scatter plots are shown in Fig. \ref{fig:fig2d}, and the regression parameters are listed in Table \ref{tab:tabparams}. Performing the partial regression analysis with these relations leads to a more significant anticorrelation: 
\begin{equation}
\log\left(\frac{\dot{M}_\mathrm{acc}}{\dot{M}_\mathrm{acc}(M,\tau)}\right) = 0.00 \pm 0.04 + (-0.27 \pm 0.07) \times \log\left(\frac{L_{\mathrm{X}}}{L_{\mathrm{X}}(M,\tau)}\right).
\end{equation}
Spearman's $\rho$ is $-0.19$ ($p < 0.001$), and the scatter is $0.74$. The linear correlation coefficient is $-0.20 \pm 0.05$. Performing the analysis with the evolutionary tracks from \cite{Dantona1994} leads to a 
slope of $-0.37 \pm 0.09$ with $\rho = -0.27$ ($p < 0.001$), $r = -0.26 \pm 0.06,$ and a scatter of $0.73$, while the tracks from \cite{Palla1999} yield a slope of $-0.36 \pm 0.08$ with $\rho = -0.26$ ($p < 0.001$), $r = -0.27 \pm 0.06,$ and a scatter of $0.72$. Thus, a significant anticorrelation is present for all three evolutionary models, with consistent slopes within $1\sigma$. A scatter plot of the values is shown in Fig. \ref{fig:figrestau}.

In order to reduce the influence of the uncertainties in the individual stellar parameters, we repeated the analysis procedure using the 
accretion luminosity $L_\mathrm{acc}$ instead of $\dot{M}_\mathrm{acc}$.
One obtains almost the same relation between the residual values: $L_\mathrm{acc} / L_\mathrm{acc}(M,\tau) \propto (L_{\mathrm{X}} / L_{\mathrm{X}}(M,\tau))^{-0.26 \pm 0.07}$.
Plotting the residual accretion luminosity as a function of the residual accretion rate reveals that both quantities are almost identical, as seen in Fig. \ref{fig:figlacc2}. 
However, this result must be regarded critically due to the large uncertainties of the isochronal ages \citep[e.g.,][]{Soderblom2014}. \cite{DaRio2014} suggest that the temporal decay of $M_\mathrm{acc}$ may be overestimated due to a connection between the uncertainties of the mass accretion rates and the uncertainties of the isochronal ages.

\subsection{Relation between mass accretion rate and plasma temperature}

The COUP X-ray spectra were fitted with a two-temperature thermal plasma model \citep{Getman2005}. We investigated the relation between
the higher temperature component, denoted as $T_2$, and the residual X-ray luminosities and accretion rates. 
For the analysis, we rejected an obvious outlier with $\log(T_2~[\mathrm{K}]) > 8$.

The results are summarized in Fig. \ref{fig:figplasma} and Table \ref{tab:results}. Both relations display a significant positive correlation,
independently of the chosen evolutionary model. This result suggests that the hardness of X-ray spectra, which is associated with higher plasma
temperatures, may influence the accretion rates in such a way that stars with harder X-ray spectra display higher accretion rates, on average, 
for a given mass and age. The development of new X-ray photoevaporation models, performed with spectra of different hardnesses, is in progress (Ercolano et al., in preparation).
Since stars with harder X-ray spectra also have a higher X-ray luminosity on average, this effect counteracts the expected anticorrelation between X-ray luminosities
and mass accretion rates. This mechanism may provide at least a partial explanation for the weakness of the observed anticorrelation, alongside 
the various uncertainties of the concerned parameters.


\section{Summary and conclusions}
We have presented a new analysis of X-ray luminosities of young stars in the ONC determined with \textit{Chandra} as part of the COUP survey
and with accretion data obtained from the photometric catalog of the \textit{HST Treasury Program}.
We have found a weak anticorrelation between residual X-ray luminosities and stellar accretion rates in a sample of $332$ young stars. 
Since all relations analyzed in this work show a large scatter, a strong correlation could not be expected. Furthermore, it cannot
be expected that the correlation becomes stronger using a larger sample. The nature of the large scatter is not fully understood. Apart from observational uncertainties, the stellar parameters are also plagued by intrinsic uncertainties due to the physical nature of young stars (see the discussion in \cite{DaRio2014}). For instance, the X-ray luminosities likely suffer intrinsic differences in the X-ray activity levels of the YSOs \citep{Preibisch2005}. Both the X-ray luminosities and the 
accretion rates show a temporal variability that contributes to the observed scatter \citep[e.g.,][]{Getman2005,Venuti2015}. A detailed
analysis of the temporal variability of the accretion rates for the young stars in the ONC and its influence on the relation between X-ray
emission and accretion is in progress (Flaischlen et. al, in preparation).

Our analysis showed that a stronger and more significant anticorrelation is obtained when the X-ray 
luminosity calculation is restricted to the soft X-ray regime ($0.5 - 2.0~\mathrm{keV}$) or when strong flares are excluded. 
Taking the isochronal age dependence of the parameters into account, the anticorrelation
is even more significant. Furthermore, we have found evidence that stars with harder X-ray spectra 
have higher X-ray luminosities and higher accretion rates on average.
Both correlations are predicted by theoretical models of X-ray photoevaporation \cite[Ercolano et al. in preparation]{Ercolano2008a, Ercolano2009, Owen2010, Picogna2019} and support the idea that protoplanetary disks undergo a phase of photoevaporation-starved accretion, as suggested by \cite{Drake2009}.
 
\begin{acknowledgements}
We wish to thank the referee for helpful suggestions.
This work was funded by the \emph{Deut\-sche For\-schungs\-ge\-mein\-schaft}
(DFG, German Research Foundation) in project PR 569/14-1 in the context of the
Research Unit FOR 2634/1: ``Planet Formation Witnesses and Probes:
TRANSITION DISKS''.
This project has received funding from the European Union's Horizon 2020 research and innovation programme under the Marie Sklodowska-Curie grant agreement No 823823 (DUSTBUSTERS).
This work has made use of data from the European Space Agency (ESA) mission
{\it Gaia} (\url{https://www.cosmos.esa.int/gaia}), processed by the {\it Gaia}
Data Processing and Analysis Consortium (DPAC,
\url{https://www.cosmos.esa.int/web/gaia/dpac/consortium}). Funding for the DPAC
has been provided by national institutions, in particular the institutions
participating in the {\it Gaia} Multilateral Agreement.
This research has used data obtained from the \textit{Chandra} Data Archive and
the \textit{Chandra} Source Catalogue, and software provided by the \textit{Chandra}
X-ray Center (CXC) in the application packages CIAO and TARA.
This research has also made use of the SIMBAD database
and the VizieR catalog services operated at Strasbourg astronomical Data Center
(CDS).
\end{acknowledgements}

%
%

\bibliographystyle{aa} 
\bibliography{39746_final} 

\begin{appendix} 

\section{Relation between the accretion luminosity and the stellar mass and isochronal age} \label{a0}
\begin{figure}
\resizebox{\hsize}{!}
        {
                \includegraphics[width=\hsize]{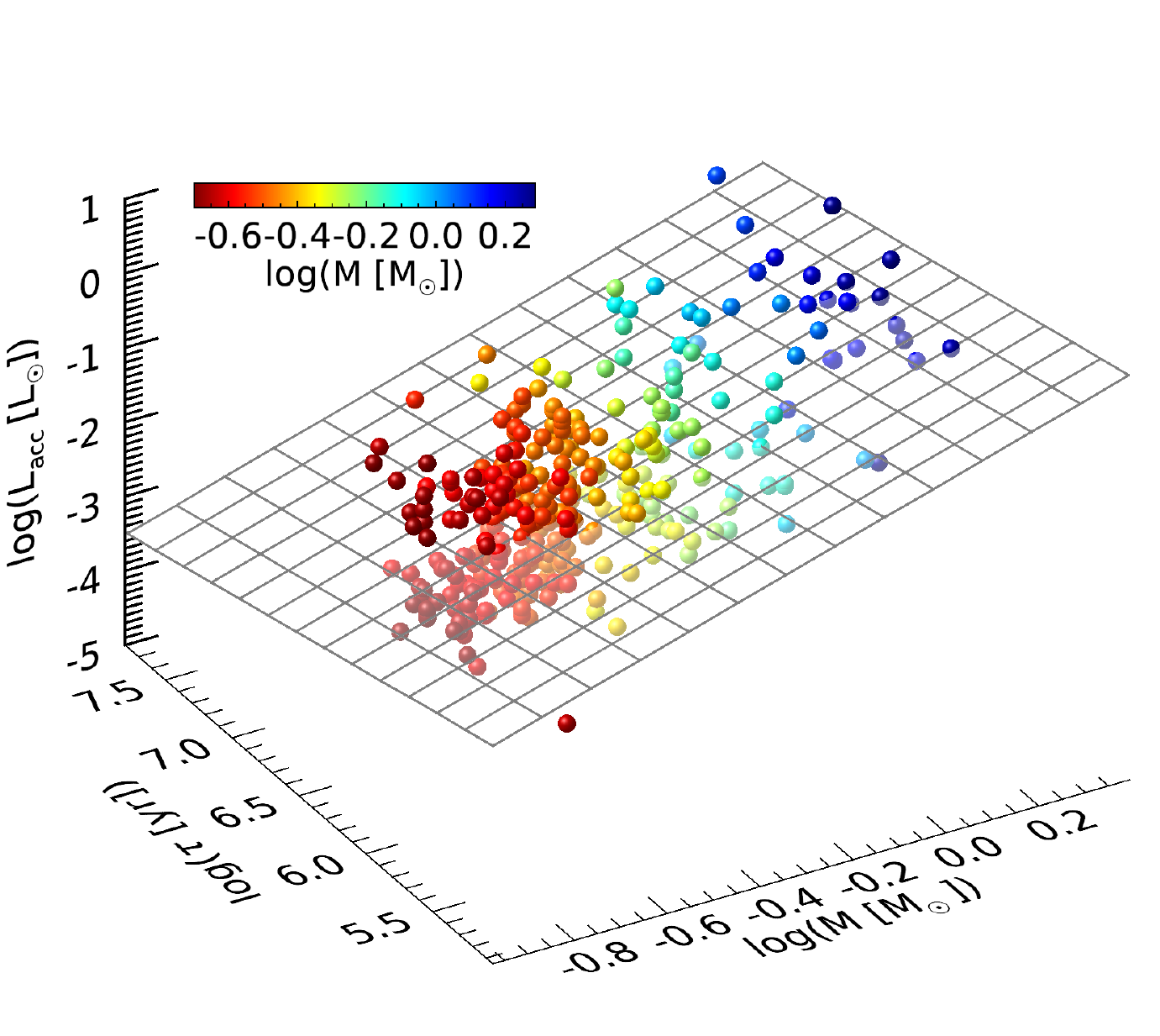} 
        }
\end{figure}

\begin{figure}
\resizebox{\hsize}{!}
        {
                \includegraphics[width=\hsize]{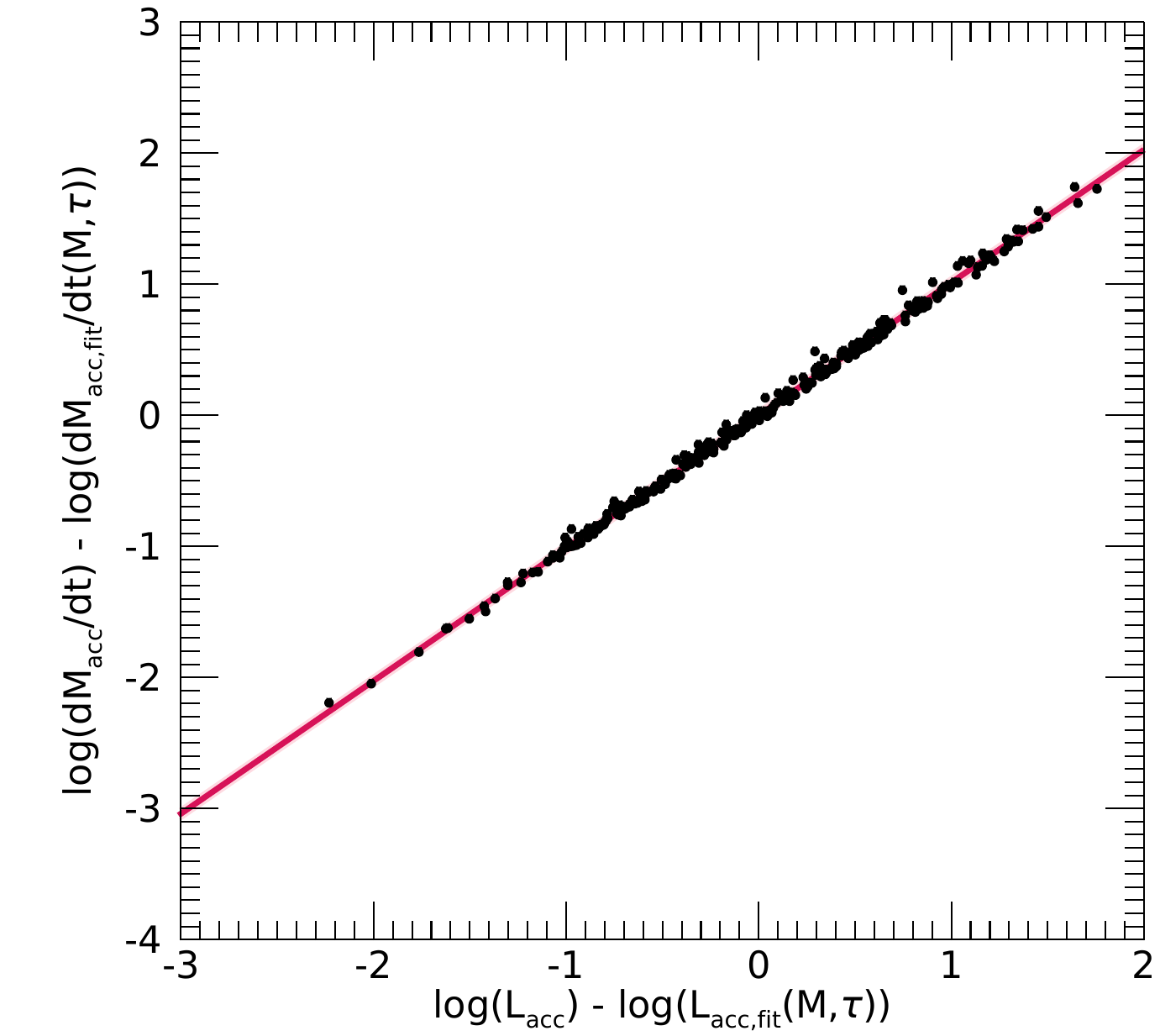}        
        }
        \caption{Correlations of the accretion luminosity $L_\mathrm{acc}$. Top: 3D plot of $L_\mathrm{acc}$ with respect to both $M$ and $\tau$. The plot points are color coded according to the masses. Bottom: Residual mass accretion rate as a function of the residual 
                        accretion luminosity.   
        } \label{fig:figlacc2}
\end{figure}

The top part of Fig. \ref{fig:figlacc2} shows a 3D plot of the accretion luminosity, $L_\mathrm{acc}$, as a function of both the mass and the isochronal age. The parameters of the fitted
plane are listed in Table \ref{tab:tabparams}. The bottom part of Fig. \ref{fig:figlacc2} shows the residual mass accretion rate as a function of the residual accretion luminosity.

\section{Method used by Telleschi and Bustamante with a synthetic data set} \label{a1}

\begin{figure}
\resizebox{\hsize}{!}
        {
                \includegraphics[width=\hsize]{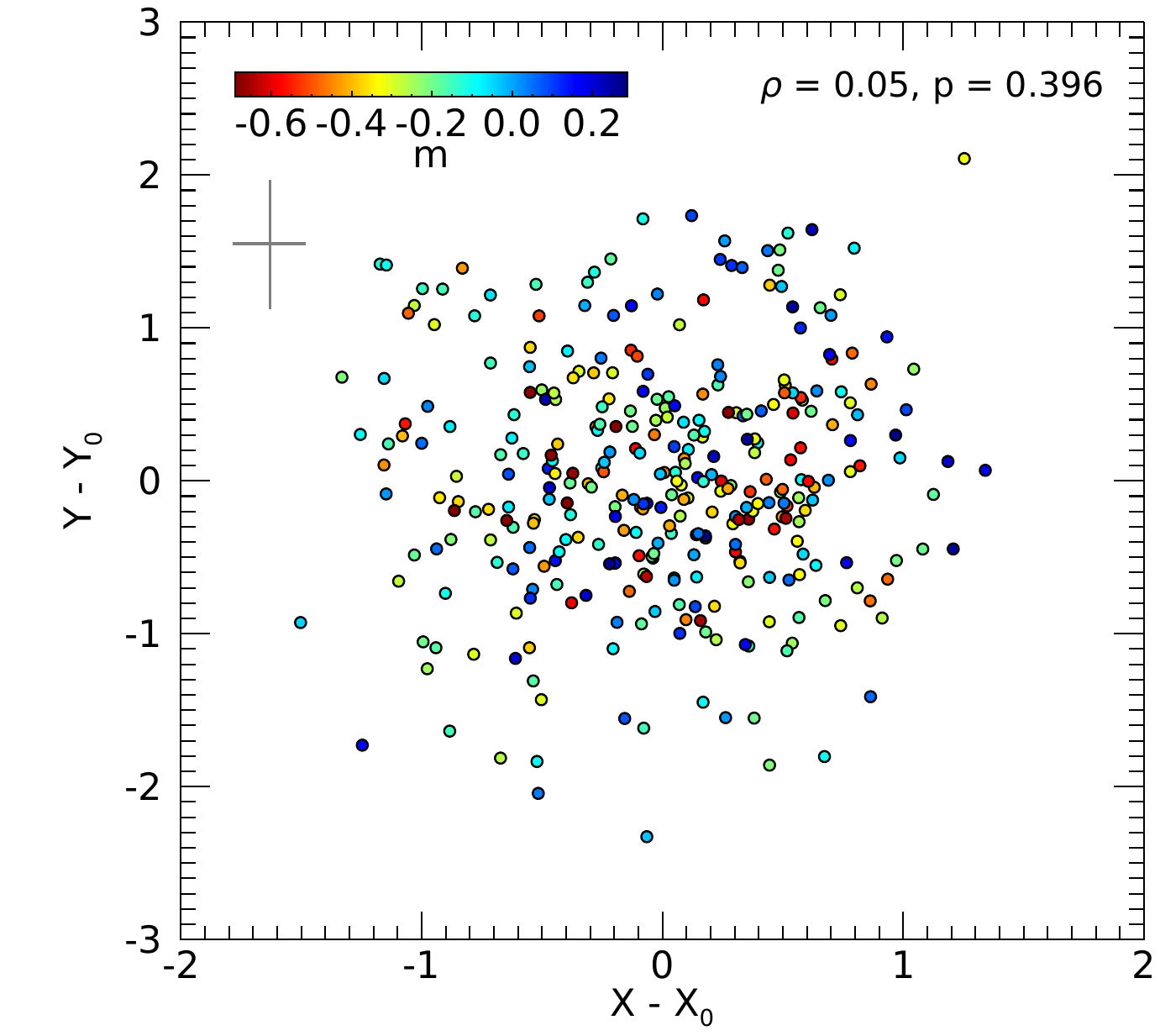}           
        }
\end{figure} 

\begin{figure}
\resizebox{\hsize}{!}
        {
                \includegraphics[width=\hsize]{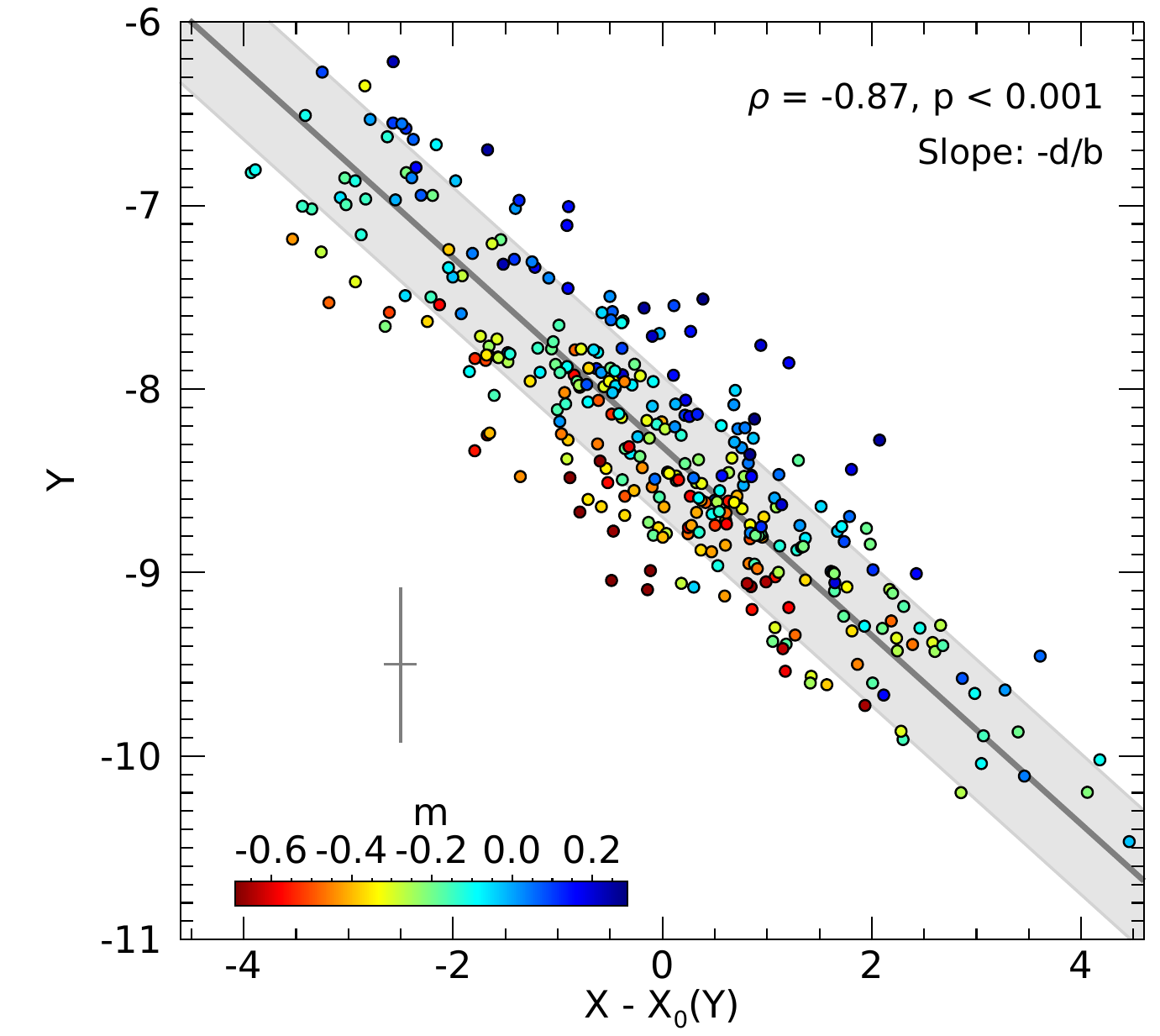}    
        }
        \caption{Correlations of the simulated data. Top: Result of the partial regression analysis on a synthetic data set with $332$ 
                        sources together with normally distributed scatter, which was chosen to be similar to the measured values.
                        There is no correlation, as expected.
                        Bottom: Method used by \cite{Telleschi2007} and \cite{Bustamante2016} applied to the same data set. The values scatter around a line 
                        with a slope of $-d / b$ and not around a constant, as assumed by \cite{Telleschi2007}. The reason is a common response 
                        variable, $s_y$, introduced as an artifact of the analysis method as described in Sect. \ref{sec:macclx}.
        } \label{fig:figxy}
\end{figure} 

In order to illustrate the issue with the method proposed by \cite{Telleschi2007} and described in Sect. \ref{sec:macclx}, we tested 
the method with a simple model and a synthetic data set. A total of $332$ logarithmic masses were generated randomly, and the values $X_0$ and $Y_0$ were 
calculated according to Eqs. \ref{eq:eqX0} and \ref{eq:eqY0}, with the parameters $a = 30.58 \pm 0.07$, $b = 2.08 \pm 0.16$, $c = -8.11 \pm 0.10$, and $d = 1.07 \pm 0.22$. The standard deviations of $s_x$ and $s_y$ were distributed normally and chosen to 
be similar to the measured values of the ONC ($0.58$ for $X_0$ and $0.80$ for $Y_0$). In this controlled model scenario, we know for sure that there is no intrinsic relation between $X$ and $Y$.
The top part of Fig. \ref{fig:figxy} shows the result of the partial regression analysis. As expected, no correlation is indicated. The bottom part shows the result of the method used by \cite{Telleschi2007} and \cite{Bustamante2016} on the same data set. One can clearly see an anticorrelation. The values scatter around a line with a slope of $-d/b$ and not around a constant. Spearman's $\rho$ suggests a strong anticorrelation as well. The color coding reveals a systematic shift in the $Y$ direction that depends on $m$. We therefore recommend not using this method to study correlations.

\begin{figure}
\resizebox{\hsize}{!}
        {
                \includegraphics[width=\hsize]{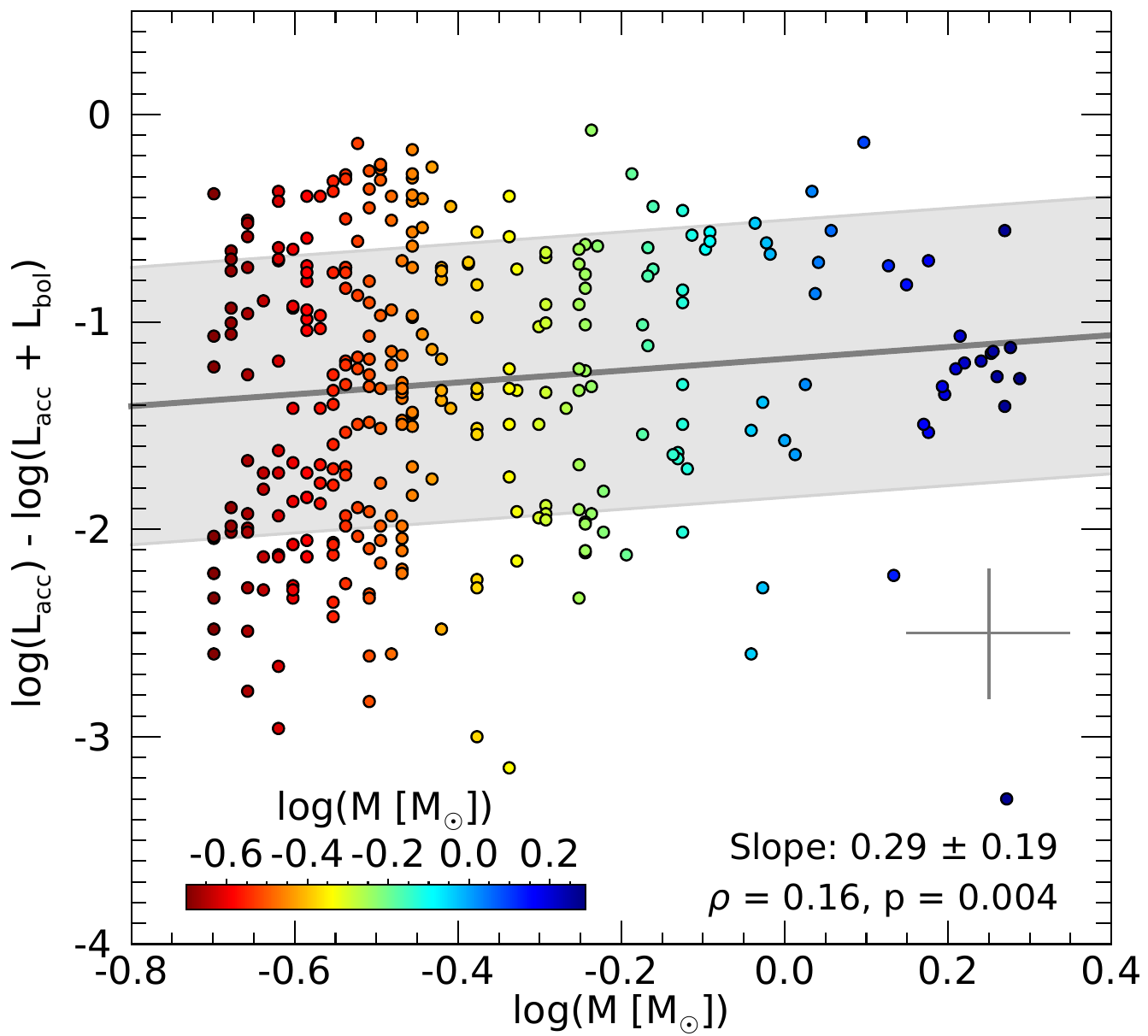}
        }
        \caption{Ratio $L_\mathrm{acc} / (L_\mathrm{acc} + L_\mathrm{bol})$ as a function of the stellar mass. The cross indicates the typical uncertainties. There is a weak positive correlation.
        } \label{fig:ratio}
\end{figure} 

\section{Proportion of the accretion luminosity on the total luminosity} \label{sec:prop}
Figure \ref{fig:ratio} shows the relation between the logarithmic mass and the logarithm of the ratio 
$L_\mathrm{acc} / (L_\mathrm{acc} + L_\mathrm{bol})$. The typical uncertainty of $L_\mathrm{acc}$ was estimated to be $0.25~\mathrm{dex}$ \citep{Alcala2017}. We assumed a similar value for $L_\mathrm{bol}$ and propagated the error to the ratio. There is a weak positive correlation ($\rho = 0.16, p = 0.004$). A linear regression using the \verb|LINMIX_ERR| routines yields a slope of $0.29 \pm 0.19$ and a linear correlation coefficient of $0.10 \pm 0.07$. The scatter amounts to $0.67$. This result suggests that 
the proportion of the accretion luminosity on the total luminosity is higher for more massive stars on average.
We note that this behavior also follows from the known $L_\mathrm{acc} - M$ and $L_\mathrm{acc} - L$ relations. Given that the
latter is described by $L_\mathrm{acc} \propto L^a$ and the former by $L_\mathrm{acc} \propto M^b$, the $L_\mathrm{acc} / (L_\mathrm{acc} + L_\mathrm{bol}) - M$ relation steadily increases for $a > 1$ and $b > 0$.

\end{appendix}

\end{document}